\documentclass[journal]{IEEEtran}
\usepackage[utf8]{inputenc}
\usepackage[english]{babel}
\usepackage[autostyle]{csquotes}
\MakeOuterQuote{"}

\usepackage{amsmath,amssymb,amsfonts}
\usepackage{algorithmic}
\usepackage{graphicx}
\usepackage{textcomp}
\usepackage[table]{xcolor}
\usepackage{listings}
\usepackage{hyperref}
\usepackage{textcomp}
\usepackage{booktabs}

\usepackage[
backend=biber,
style=numeric,
sorting=none
]{biblatex}
\begin{document}

% use short links for URLs in the bibliography
\DeclareFieldFormat{url}{\href{#1}{\mkbibacro{URL}}}

\title{Dissecting the software-based measurement of CPU energy consumption: a comparative analysis}

\author{
    Guillaume Raffin*\textsuperscript{\textdagger}, Denis Trystram* \\
    *Univ. Grenoble Alpes, CNRS, Inria, Grenoble INP, LIG, 38000 Grenoble, France \\
    \textsuperscript{\textdagger} Bull SAS (Eviden, Atos group), France \\
    \textit{\{guillaume.raffin, denis.trystram\}@univ-grenoble-alpes.fr}
}

\maketitle

\begin{abstract}
Every day, we experience the effects of the global warming: extreme weather events, major forest fires, storms, global warming, etc.
The scientific community acknowledges that this crisis is a consequence of human activities where Information and Communications Technologies (ICT) are an increasingly important contributor.

Computer scientists need tools for measuring the footprint of the code they produce and for optimizing it. Running Average Power Limit (RAPL) is a low-level interface designed by Intel that provides a measure of the energy consumption of a CPU (and more) without the need for additional hardware. Since 2017, it is available on most computing devices, including non-Intel devices such as AMD processors.
More and more people are using RAPL for energy measurement, mostly like a black box without deep knowledge of its behavior.
Unfortunately, this causes mistakes when implementing measurement tools.

In this paper, we propose to come back to the basic mechanisms that allow to use RAPL measurements and present a critical analysis of their operations. In addition to long-established mechanisms, we explore the suitability of the recent eBPF technology (formerly and abbreviation for extended Berkeley Packet Filter) for working with RAPL.
For each mechanism, we release an implementation in Rust that avoids the pitfalls we detected in existing tools, improving correctness, timing accuracy and performance. 
These new implementations have desirable properties for monitoring and profiling parallel applications.

We also provide an experimental study with multiple benchmarks and processor models (Intel and AMD) in order to evaluate the efficiency of the various mechanisms and their impact on parallel software.
These experiments show that no mechanism provides a significant performance advantage over the others. However, they differ significantly in terms of ease-of-use and resiliency.

We believe that this work will help the community to develop correct, resilient and lightweight measurement tools.

\end{abstract}

\begin{IEEEkeywords}
% basés sur le dictionnaire des keywords IEEE
energy consumption, energy efficiency, performance analysis, software measurement, RAPL library (Running Average Power Limit)
\end{IEEEkeywords}

%les outils basés sur RAPL ont un overhead ! la solution proposée est meilleure que les sondes logicielles existantes...

% Les conclusions :
% - pas de différence significative sur l'efficacité,
% mais il y a des différences sur les aspects résilience, facilité d'utilisation
% - Attention, utiliser RAPL n'est pas si simple, il y a une bonne façon de faire (vis à vis de la facilité)
%
% un message final ? On a un choix à faire dans le compromis entre les qualité de l'un ou l'autre.
% On donne une recommandation basée sur l'expérience sur plusieurs benchmarks sur 2 architectures (Intel et AMD).

% Remarques de Danilo :
% - ouverture possible sur le Edge, en tout cas la différence de conso selon la fréquence a un intéret dans ce domaine, qq watts c'est beaucoup !
% - expliquer comment bien corriger l'overflow est utile, peu d'infos fiable disponibles à ce sujet, encore moins sur chaque mécanisme

\section{Introduction}
\label{sec:introduction}

\subsection{Context and motivation}
\label{subsec:context}

The impact of ICT in the environmental crisis has unfortunately increased, and it is likely going to continue increasing~\cite{malmodinFootprint2018,AndraePredictionFor2030,BelkhirElmeligiPredictionFor2040,Bol2021MooreLawAnthropocene,shiftProject2023Trends}. The carbon footprint of the field was evaluated to about 1.2 to 2.2 Gigatons equivalent CO2 (denoted in short by GtCO2e) in 2020, which corresponds to about 2.1 to 3.9\% of worldwide greenhouse gases (GHG) emission~\cite{Freitag}. They are expected to reach 5.1 to 5.3 GtCO2e by 2040~\cite{BelkhirElmeligiPredictionFor2040}.
Yet, ICT are also often presented as an effective solution for decreasing the environmental footprint of other sectors, mostly thanks to optimization and substitution effects~\cite{Hilty2006,Hilty2015}.
Some reports hence claim that ICT's environmental benefits can be several times greater than their own environmental burden~\cite{aiTechnophile}.
Nevertheless, the assessment of these benefits generates a lot of controversy~\cite{Freitag,Rasoldier2022,coroamaMethodo}, as the global pressure of humanity on the environment keeps on increasing~\cite{IPCC_2022_WGIII, SteffenPlanetaryBoundaries}.
In this context, measuring the energy consumption of ICT is required to become aware of its impact (such as its carbon footprint) and establish environmentally friendly practices such as optimization efforts, power capping, restriction use, etc.
%In particular, knowing the energy consumption of a system, be it a whole supercomputer platform, a single node or a particular piece of software, allows to estimate its carbon footprint.
% ^^^ commenté car pas assez de place

Section 4 of the paper of Khan et al.~\cite{KhanButtTop500Trends} shows that energy efficiency is increasing in the Top500, but at a lower rate as computing performance. \textit{Aurora} became the second machine to reach exascale in June 2024~\cite{top500}, with a record-breaking consumption of 38.7~MW with only about 40\% of its nodes. This generates about 270 tons of CO2e emissions a day (according to Illinois 2022 electricity profile). With such an impact, HPC systems are more and more studied from the perspective of energy efficiency. Assessing the energy efficiency of a system calls for measuring the energy consumption.

% Rapport annuel 2022 de l'EIA : https://www.eia.gov/electricity/state/illinois/
% donne 639 lbs CO2 par MWh
% => 289.8kg CO2 / MWh
% => 24h * 38.7MW = 928.8 MWh
% => 928.8*289.8 = 269166 kg CO2

There are multiple families of tools that can be used to collect the energy consumption ranging from physical wattmeters (e.g. connected PDUs -- Power Distribution Units -- or BMCs -- Baseboard Management Controllers) to web browser extensions that run estimation models.
They have different accuracy, acquisition frequency and ease-of-use.

In this paper, we focus on the Running Average Power Limit (RAPL) technology~\cite{intelManual}, which is the basic building block for many software measurement tools.
RAPL measures the electricity consumption of the CPU and more components thanks to sensors integrated into the system-on-chip, and exposes it to the operating system through model-specific registers (MSR in short) designed by the CPU manufacturer. The advantage of RAPL is that no external powermeter is required, nor a privileged access to the BMC (which could be used to power off the server). It also provides more details than a PDU or a BMC, since it monitors internal components of the node.
Moreover, RAPL is more accurate than any statistical estimation model, even though they can be tuned to reduce their error~\cite{heinrich_predicting_2017, fieni_smartwatts_2020}.
The underlying philosophy of software measurement tools based on RAPL is that each device owner monitors the software running on it.

Extracting the measurements from the CPU requires an interface. One can either use the low-level RAPL MSR directly, or choose a higher-level interface provided by the operating system. Linux provides two of them, namely, the Power Capping framework (powercap) and the Performance Counters subsystem (perf-events). These software interfaces query the processor registers provided by RAPL.

Sometimes, the same interface can be used in a number of very different ways. This is the case with perf-events, which can be read from user space or from kernel space using eBPF. We propose to study these two ways separately. This makes a total of four mechanisms that can be chosen to extract the consumption measurements from the CPU: MSR, powercap, perf-events in user space, and perf-events with eBPF. They are described in section~\ref{sec:RAPLmechanisms}.

Higher level software measurement tools, such as CodeCarbon~\cite{CodeCarbon}, PowerAPI~\cite{bourdon2013powerapi} and Scaphandre~\cite{Scaphandre}, are based on these raw mechanisms. They use RAPL measurements to estimate the electricity consumption of each active process.

\subsection{Objectives}
\label{subsec:objectives}

The purpose of this work is to provide engineers and researchers that use or want to use RAPL with a deep understanding of its access mechanisms and of the associated best practices. From their point of view, the goal is to develop measurement tools that are correct, efficient, and maintainable over time.

In order to achieve this goal, we need to evaluate the qualitative differences between the possible measurement mechanisms, based on criteria such as required expertise level and resiliency. This highlights the features and trade-offs of the mechanisms. Furthermore, we assess the overhead of the various mechanisms. In particular, we address the following questions:

What is the overhead on the other programs of measuring the energy consumption of the CPU? What is the impact of the measurement on an idle machine? Are some mechanisms more efficient than others? Is there a difference between Intel and AMD processors? To the best of our knowledge, these issues have not been fully addressed yet.

\subsection{Contributions}
\label{subsec:contributions}

To achieve the previous objectives, we propose a minimal measurement tool that includes an implementation of each mechanism. We release it with an open-source license in order to provide a reference implementation to the community. We test different strategies to avoid the common mistakes that we detected in other tools. Furthermore, we analyze and compare the four measurement mechanisms.

Using this minimal tool, we experimentally study the impact of the measurement on parallel software and on an idle server. Each mechanism has been benchmarked with several HPC testbeds, CPU vendors and RAPL domains. The benchmark results have been accumulated during approximately one month, with the aim to reduce the uncertainty of the statistical analysis.

In light of the comparative analysis and of the benchmark results, we are able to give a recommendation on the choice of the right mechanism.

%%%%%%%%%%%%%%%%%%%%%%%
\section{Related Works}
\label{sec:relatedWorks}

% TODO ajouter kocotEnergyAwareHPC ? (broad study on available technologies, no deep dive)
%
% focus on hardware sensors: burtscher_measuring_2014, almeida_energy_2018, laros_powerinsight_2013
%
% estimation models: smartwatts, heinrich_predicting_2017
%
% accuracy (en plus de ceux listés) : sen_quality_2018 (GPU)

Hackenberg et al. evaluated the accuracy of RAPL measurements, in 2013~\cite{hackenberg_power_2013} and 2015~\cite{hackenberg_energy_2015}, using the MSR mechanism. Measurement overhead was not considered.

Huang et al.~\cite{huangHaswellPowerMeasurements} showed that RAPL measurements were quite accurate for Haswell-EP processors. They assessed the performance overhead of the PAPI library. The different low-level mechanisms were not investigated.

Descrochers et al.~\cite{desrochers_validation_2016} compared the values returned by RAPL for the RAM with power measurements obtained by instrumenting the hardware. Like the previous papers, this one concentrated on RAPL accuracy, not on its overhead.

In 2018, Khan et al.~\cite{raplInAction} evaluated the accuracy, update frequency and performance overhead of RAPL measurements on several parallel benchmarks. Their work focused on the MSR mechanism and did not compare it with others.

Several works were dedicated to the construction of statistical models that estimate the energy consumption of software applications~\cite{fieni_smartwatts_2020,heinrich_predicting_2017,garcia-martin_estimation_2019}. While they compared their results with RAPL measurements, investing the various mechanisms was out of their scope.

In 2021, Schole et al. analyzed the measurements of AMD's implementation of RAPL on the Zen 2 architecture, and revealed that it provided less information than Intel's~\cite{schone_energy_2021}. One mechanism was operated, and the study concentrated on assessing the reliability of AMD's implementation.

More recently, Jay et al.~\cite{JayCcgrid} compared some high-level tools that internally use RAPL. They evaluated their accuracy and overhead on NAS parallel benchmarks. We also use the NAS benchmarks, and conduct a more robust statistical analysis of their results.
A similar study was proposed by Heguerte et al. for AI models~\cite{LuciaEstimateFootprintDeepLearning}. We go deeper than both papers and analyze the underlying mechanisms instead of the higher level tools.

Due to the amount of work on RAPL measurements validation, we consider that this technology is reliable enough. Therefore, we choose not to assess again the correlation between RAPL values and external measurements. Instead, we focus on a comparative analysis of the measurement mechanisms, based on qualitative criteria as well as an experimental evaluation of their performance and energy overhead.

%%%%%%%%%%%%%%%%%%%%%%%%%%%%%%%%%%%%%%%%%%%%%%%%%%%%%%%%%%%%
\section{Limitations of the existing approaches}
\label{sec:Problems}

To understand why it is relevant to open the black box of RAPL, let us begin by an overview of the pitfalls that we have detected in many existing RAPL-based tools. We believe that these limitations result from the lack of a precise understanding of RAPL and the interfaces provided by the Linux kernel on top of it.

\subsection{Unclear measurement perimeter}
\label{subsec:problems-perimeter}

RAPL, that stands for Running Average Power Limit, is a power management technology first implemented by Intel in the Sandy Bridge architecture, in 2011. It allows to measure the energy consumption and to limit the power consumption of various parts of the computing device, called domains. In this paper, we only look at the measurement interface, not the power limit. AMD followed suit by implementing a similar measurement interface in the first Zen architecture, in 2017. AMD's interface is so similar that it is used in the same way as Intel's. The difference lies in the addresses of the registers and in the number of available domains. In both cases, understanding what is measured by each domain is crucial.

Unfortunately, the scopes of the domains are not clearly explained in the documentation of the Linux kernel interfaces that are used by the measurement tools. The Intel Software Developer Manual~\cite{intelManual} provides more information, but it remains vague about the most recent domains such as \textit{"psys"}. Consequently, several contradictory descriptions of the domains can be found online, and it is difficult to understand what is really measured.

\begin{figure}[h]
    \centering
    \includegraphics[width=\columnwidth]{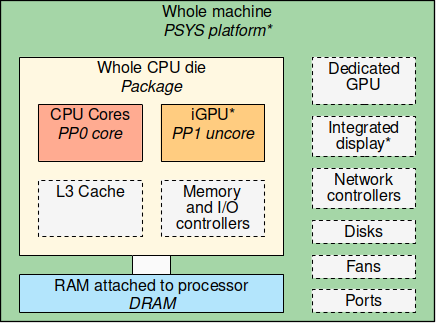}
    \caption{Hierarchy of the possible RAPL domains and their corresponding hardware components. Domain names are in italic, and grayed items do not form a domain on their own. Items marked with an asterisk are not present on servers.}
    \label{fig:raplDomains}
\end{figure}

By completing the documentation with experiments, we offer a new representation of all the RAPL domains known to date, which can be seen in figure~\ref{fig:raplDomains}. Khan et al.~\cite{raplInAction} published a similar figure, but we provide here some clarifications and corrections, especially with regard to the role of \textit{"psys"}.

First, figure~\ref{fig:raplDomains} shows that the \textit{core} and \textit{uncore} domains are subsets of the \textit{package} domain.
This means that their reported consumption is included into the consumption of the \textit{package} domain.
Second, let us take a look at the \textit{platform} domain, also known as \textit{psys}. According to Intel~\cite{intelPsysPatch}, its content depends on the manufacturer of the machine. To understand its actual scope, we used recent laptops (Lenovo Thinkpad L15 Gen1 and Gen2, and Alienware m17 R3) and plugged each of them into an external wattmeter. By comparing the energy consumption reported by the wattmeter with the consumption reported by the RAPL domains, we discovered that the \textit{psys} domain reported the same consumption as an external wattmeter, that is, the total consumption of the laptop. This domain can therefore include the display, dedicated GPU and all the other domains.
Third, we want to make clear that, while the \textit{psys} domain is unique (i.e. there is only one \textit{psys} for the entire system-on-chip), the \textit{dram} domain is linked to a specific CPU socket (i.e. there is one \textit{dram} per socket). This clarifies what the Intel manual means by "directly-attached RAM". We have tested this using multi-CPU servers Gigabyte H261-Z60.
Finally, AMD's microarchitectures Zen 1 to Zen 4 only support the \textit{package} and \textit{core} domains.

\subsection{Overflows that distort measurements}
\label{subsec:problems-overflows}

Measurements provided by RAPL have been evaluated by prior work to be accurate. However, bugs in measurement software can distort this accuracy. In particular, a potential difficulty of RAPL is that the energy counters overflow. These overflows occur after approximately "60 seconds under heavy load" according to Intel's manual~\cite{intelManual}, and the only way to detect them is to poll the value frequently enough (the CPU does not signal them). It may appear as a trivial bug, however in this case it can actually be hard to detect, because the overflow frequency depends on the consumption of the machine.
Testing the program on a mainstream laptop or on a server that is mostly idle can therefore make it invisible to the tester's eyes. Furthermore, if the user look at aggregated indicators, detecting the error becomes even harder.

We have found that many software measurement tools have been affected by bugs related to this overflow. Not all tools have resolved the issue as of this writing~\cite{bugScaphandre, bugCodeCarbon, bugLikwid}, which could cast doubt on the measurements they provide. A robust solution is presented is section~\ref{subsec:overflowCorrection}.

\subsection{Lack of accurate control of the acquisition frequency}
\label{subsec:problems-frequency}

As stressed previously, the acquisition frequency is a key point of a RAPL-based measurement tool. Besides the overflow correction, one could wish to increase the frequency in order to capture a more precise profile of the running applications. For example, estimating the consumption of a single function has been proven to be possible in 2012~\cite{hahnel_measuring_2012}. To do so, we have to know the beginning and end of the function, and to maintain an acquisition frequency that is high enough to distinguish the different function calls, but low enough to minimize the disruption of the system. Parallel applications, which use multiple cores and call complex libraries, require even more caution.
For these reasons, it is necessary to give control of the frequency to the end user, and to make the tool precisely follow the supplied frequency.

Unfortunately, this is not the case with the measurement tools that we tested. Usually, the acquisition frequency is capped to 1~Hz, and the actual frequency differs from the target. We found that the implementation of the polling loop had a great influence on the reliability of its timing. In particular, using a traditional \texttt{sleep} function (like Scaphandre~\cite{Scaphandre}, PyJoules~\cite{inria_spirals_pyjoules_2019} and others) proved to be unreliable: as assessed by our experiments, the time between each measurement varies and the high frequencies cannot be achieved. Thus, we believe that this approach cannot be used for energy profiling. We were able to overcome this issue by implementing a minimal tool using a different technique, which is described in section~\ref{subsec:ensuringAccuracy}.

\subsection{Performance and quality concerns}
\label{subsec:problems-perf-quality}

Every tool that we tested only supports one RAPL-based measurement mechanism, but does not justify its choice. Some tools like Likwid~\cite{gruber_likwid_2021} use an MSR-based mechanism, while others like CodeCarbon~\cite{CodeCarbon} use a powercap-based mechanism. Most of the works presented in section~\ref{sec:relatedWorks} are based on the MSR. Using the low-level MSR seems harder a priori, but could have been chosen to obtain a performance advantage. We show that there is actually no such advantage in section~\ref{subsec:benchmark-results}.

The choice of the mechanism also raises a quality concern. First, it impacts the amount of work required to create the tool and maintain it over time. We analyze the qualitative aspects of RAPL-based mechanisms in section~\ref{sec:RAPLmechanisms}. Second, while implementing our minimal measurement tool, we found that bugs in the Linux kernel could impede the operation of the interfaces. On the AMD server used to conduct our benchmarks, powercap and perf-events did not list the same available RAPL domains. Powercap listed two domains: \textit{package} and \textit{core}, whereas perf-events listed all the possible domains. It turned out that only the \textit{package} domain made sense for this CPU model. The bug has been fixed in Linux kernel version 6.6~\cite{bugPerfEventsAMD}, but datacenters take time to update their systems. Supporting multiple mechanisms would allow such bugs to be detected automatically, making the tools more robust.

% A manual check of each domain revealed that both lists were probably wrong, because most of the domains were unusable, and the \textit{core} domain reported extremely low values of about $0.001$ Joule per second, which we found dubious.

\section{Implementation of a minimal tool}
\label{sec:impl-tool}

\subsection{Architecture}
\label{subsec:impl-tool-arch}

\begin{figure}[h]
    \centering
    \includegraphics[width=\linewidth]{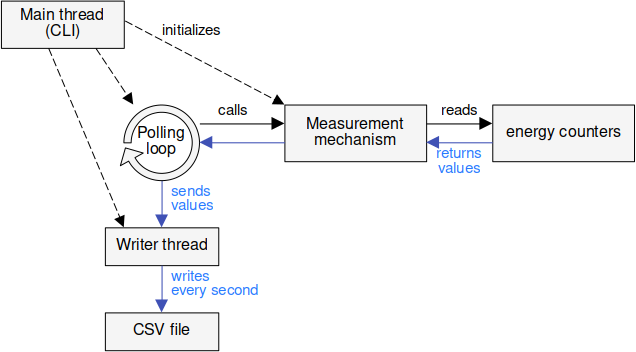}
    \caption{Architecture of our measurement tool}
    \label{fig:ourToolArch}
\end{figure}

To evaluate some solutions to the aforementioned issues and compare the four RAPL-based mechanisms, we implemented a minimal measurement tool in Rust. Its architecture is depicted on figure~\ref{fig:ourToolArch}.

It features a command-line interface that allows the user to choose the domains to measure, the mechanism to use, the acquisition frequency of the polling loop, and the output file. We chose to always write the measurements during our benchmarks, in the CSV format, in order to be more representative of a real situation. Indeed, discarding the values as soon as we read them would be useless. Thus, the impact that we evaluate in section~\ref{sec:experiments} includes the cost of saving the measurements.
To limit this cost when the acquisition frequency is high, while regularly making the data available to the user, we chose to flush the measurements to the output file once per second.

The minimal tool allows to access the "raw" energy consumption measurements, not a derived metric like Code Carbon~\cite{CodeCarbon}. We believe that this is more rigorous and that it gives more freedom to use the measurements in a way that is suited to the context. Furthermore, it allows to ensure that the overflow correction presented in section~\ref{subsec:overflowCorrection} is correct.

More information is available on a public Git repository~\footnote{https://github.com/TheElectronWill/cpu-energy-consumption-comparative-analysis: minimal tool with an implementation of each studied mechanism}.

\subsection{Correcting the overflow of the counters}
\label{subsec:overflowCorrection}

As mentioned in section~\ref{subsec:problems-overflows}, RAPL energy measurements are prone to overflows, whose frequency depends not only on the energy consumption of the domains but also on the measurement mechanism, because they store the values in variables of different sizes and sometimes apply transformations on them. To ensure the correctness of the measurements, two conditions must be met.
First, the time between two readings of the energy consumption must be smaller than the time it takes for an overflow to occur. This allows to detect an overflow between two successive measurements $m_{prev}$ and $m_{current}$, as highlighted by previous work on RAPL.
Second, the following correction must be applied:

$$
\Delta m =
\begin{cases}
    m_{current} - m_{prev} + C &\text{if}\ m_{current} < m_{prev} \\
  m_{current} - m_{prev} &\text{otherwise} \\
\end{cases}
$$

where $C$ is a correction constant that depends on the chosen mechanism. Table~\ref{table:overflowCorrection} gives the right value of $C$ for each mechanism. Only then can $\Delta m$ be used as a measure of the energy consumption of the RAPL domain during the small time period between the two measurements.

\begin{table}[h!]
    \centering
    \begin{tabular}{|m{0.15\linewidth} | m{0.75\linewidth}|}
     \hline
     mechanism & constant $C$\\ [0.5ex]
     \hline
     MSR & \texttt{u32::MAX} i.e. $2^{32}-1$ \\
     \hline
     perf-events & \texttt{u64::MAX} i.e. $2^{64}-1$ \\
     \hline
     perf-events with eBPF & \texttt{u64::MAX} i.e. $2^{64}-1$ \\
     \hline
     powercap & value given by the file \texttt{max\_energy\_uj} in the sysfs folder of the RAPL domain \\
     \hline
    \end{tabular}
    \vspace{.5em}
    \caption{Table of the overflow correction constant for each measurement mechanism}
    \label{table:overflowCorrection}
\end{table}
\vspace{-1cm}

\subsection{Ensuring the accuracy of the acquisition frequency}
\label{subsec:ensuringAccuracy}

When we implemented the tool, we found that using the \texttt{sleep} function in the polling loop was not reliable enough. The variations between each sleep were big enough to be detectable, and it was impossible to reliably reach a frequency of 1000~Hz. To solve this issue, we replaced the standard sleep by \texttt{timerfd}. While it cannot guarantee to perfectly respect the frequency in all conditions, especially on a non-realtime kernel, we assessed its superiority over a standard sleep.

In addition to using a more precise sleep function, we designed the tool so that the polling loop only does the minimum amount of work required to gather the measurements. Writing the data to a file is delegated to another thread. This way, the polling loop is not impacted by I/O latency.

%todo REPLACE OLD EXPERIMENTS by NEW ONES + explain new experiments
%As highlighted in section~\ref{subsec:problems-frequency}

To evaluate the effect of these two optimizations, we built two versions of the minimal tool: one that included the optimizations, and one that did not. We executed them on an idle laptop, gradually increased the target frequency, and computed the actual output rate, i.e. the number of measurements per second in the result file. The first and last second were removed from the data, in order not to skew the statistics with incomplete measurement periods.

\begin{figure}[h]
    \centering
    \includegraphics[width=\linewidth]{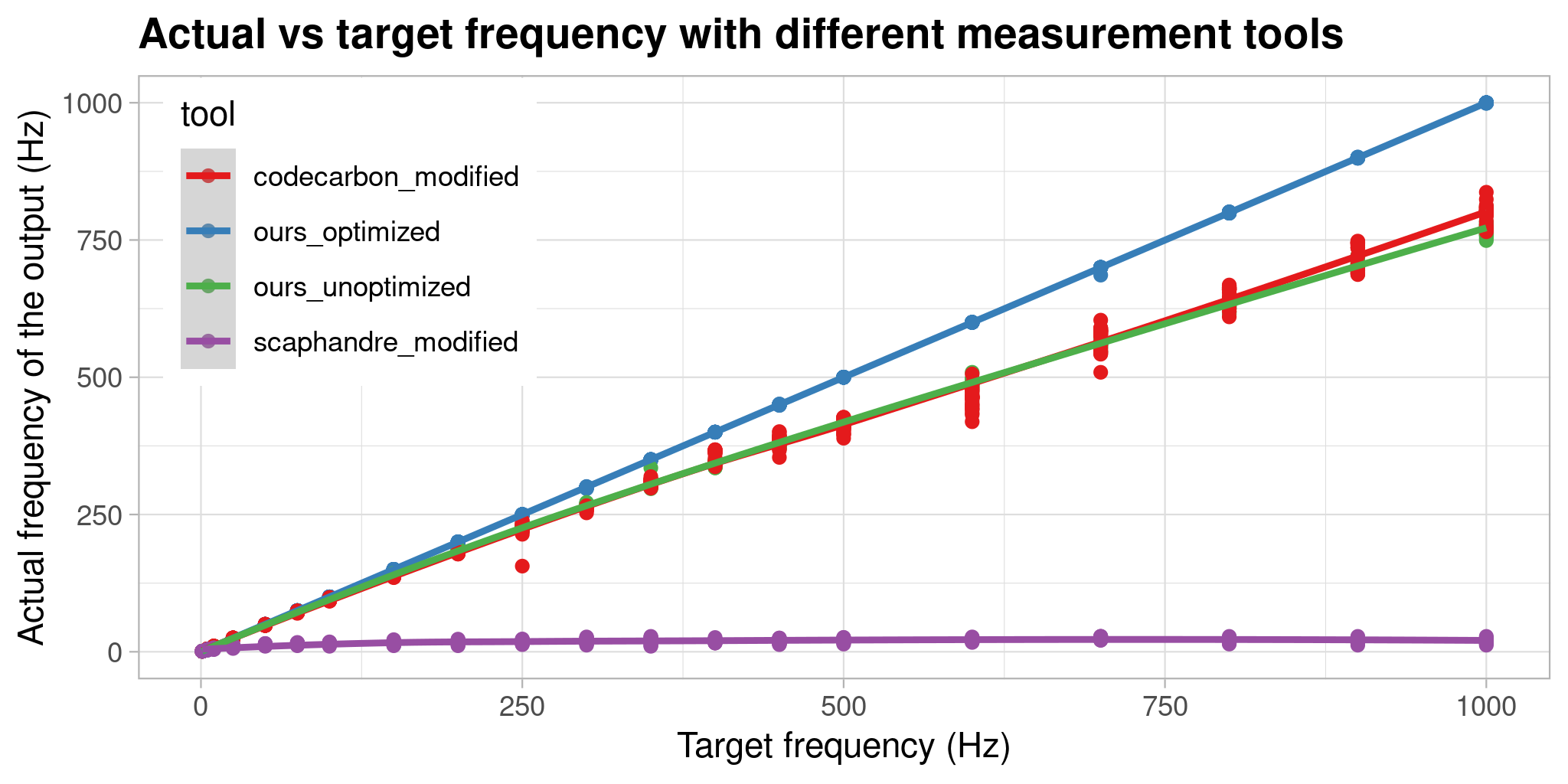}
    \caption{Plot of the actual measurement frequency achieved by the tools for each target frequency supplied as a command-line argument (on a Lenovo Thinkpad L15 Gen1 laptop)}
    \label{fig:measurementFreqPlot}
\end{figure}

We did the same experiment with two other tools: Scaphandre~\cite{Scaphandre} and CodeCarbon~\cite{CodeCarbon}. We chose them because they are among the most popular open-source RAPL-based tools, and because they differ in some key implementation choices. In particular, Scaphandre~\cite{Scaphandre} uses a \texttt{sleep}-based timing loop and CodeCarbon~\cite{CodeCarbon} uses a timer library for Python. We modified them to allow measurement frequencies above 1~Hz. To keep the comparison fair, we also disabled a part of their code in order to only collect the raw RAPL measurements. Their measurement perimeter was thus identical to our minimal tool. Figure~\ref{fig:measurementFreqPlot} shows the results obtained by each measurement tool on the laptop.

% \begin{figure}[h]
%     \centering
%     \includegraphics[width=\linewidth]{ggplot/timing-accuracy-boxplot.png}
%     \caption{Boxplot of the number of measurements per domain per second, for each tested equipment (node) and version: fully optimized, without timerfd (badsleep) and withour timerfd nor the I/O thread (badsleep-st)}
%     \label{fig:boxplotTimingAccuracy}
% \end{figure}

As shown on the plot above, only the optimized version of the minimal tool (in blue) is able to consistently reach the target frequency of 1000 Hz with a very low standard deviation of 0.7. With the unoptimized version, the actual frequency decreases to about 847 Hz (15\% lower than the target) and the standard deviation increases to about 20. We also tested that enabling only one of the two optimizations is not enough to obtain the best results. This demonstrates the benefit of combining them in an efficient measurement loop.

A surprising result is that, while CodeCarbon is written in Python and Scaphandre is written in Rust, the former performs significantly better than the latter. The modified version of CodeCarbon exhibits a frequency shift that is similar to our unoptimized tool, whereas the modified version of Scaphandre is unable to go beyond 28 Hz. This probably indicates that the implementation of Scaphandre is suboptimal.

CodeCarbon, despite using a timer-based loop, does not perform better than our unoptimized tool, which is sleep-based. Furthermore, it consumes significantly more CPU time by constantly using about 37\% of one core at 1000~Hz, whereas our unoptimized tool uses only 8\% at most.

We replicated the experiments on a server with two Intel Xeon Gold 6130 CPUs and obtained similar results. A notable difference was the improvement caused by the separate I/O thread, which we found to be smaller on the server than on the laptop. This can be explained by the size of the OS disk cache, which is larger on the server.

Overall, the experiments show that the two simple optimizations of the minimal measurement tool have made it accurate, and allowed it to achieve a high measurement frequency. This is a clear improvement over existing RAPL-based tools, even when the latter are modified to allow high frequencies.

\subsection{Brief discussion about limitations}

Although it allowed us to perform extensive experiments, the minimal tool that we built is limited in some ways.

First, it does not support the hot-plugging of CPUs, because this feature was useless for our experiments. It could be added by using the Linux hotplug API to be notified when a CPU goes offline or online~\cite{linuxHotplug}. % https://www.kernel.org/doc/html/v4.13/core-api/cpu_hotplug.html#using-the-hotplug-api

In addition, our work is strictly limited to Linux. Although the MSR should be accessible from other operating systems, we have not tried to use them. The low-level access of the registers would be similar but the higher-level interfaces, if any, would be different. To the best of our knowledge, Windows does not provide a sysfs-like interface to access RAPL measurements.

Finally, we have focused here on x86 Intel and AMD processors. ARM processors have not been studied because, as far as we know, most of them do not provide a RAPL-like interface. It seems to be possible on some platforms, though, since the \textit{powermetrics} tool~\cite{applePowermetrics} is able to report the instantaneous power of Apple's ARM chips. Incorporating another OS and another low-level interface into the minimal tool would require more exploratory and technical work.

\section{Comparative analysis of the RAPL-based mechanisms}
\label{sec:RAPLmechanisms}

% \begin{figure}[h]
%     \centering
%     \includegraphics[width=\columnwidth]{diagrams/rapl-mechanisms.png}
%     \caption{The different available RAPL-based energy consumption recording mechanisms}
%     \label{fig:raplMechanisms}
% \end{figure}

\subsection{Comparison criteria}

Section~\ref{subsec:problems-perimeter} introduced the fundamental operation of the RAPL technology, which provides energy counters associated to domains. To read these counters on Linux, several mechanisms can be implemented. By leveraging the experience that we obtained by building the minimal measurement tool, we now analyze and compare the possible choices. Three of them are relatively well-known, yet rarely understood in depth. We also offer a new way of reading RAPL counters and compare it to the existing ones.

All the mechanisms enable the same end result: obtaining, at a given acquisition frequency and for a given set of domains, the values of the energy consumption counters provided by the CPU's RAPL interface. A measurement tool can be based on any of them, but that choice implies some trade-offs. In order to compare the mechanisms, we have retained the following criteria:
\begin{itemize}
    \item the \textbf{technical difficulty} required for setting up the mechanism, i.e. the amount of time it takes to implement it, and the experience required for someone to do it.
    \item the amount of \textbf{knowledge} required to implement the mechanism properly, for instance whether we need to know some details of the CPU's microarchitecture or not. It depends on the abstraction offered by the system interface.
    \item the \textbf{safeguards} it offers to the developer, i.e. whether is it easy to make mistakes when using the mechanism
    \item the \textbf{privileges} required to execute the mechanism, e.g. whether the end user of the software tool that uses the mechanism has to be \texttt{root}
    \item the \textbf{resiliency} of the mechanism, i.e. its ability to adapt to changes, such as the release of a new generation of processors, a new RAPL domain, etc.
\end{itemize}

\subsection{Model Specific Registers}
\label{subsec:msr}

At the lowest level, the CPU provides model specific registers (MSR) exposing the RAPL data\cite{intelManual}.
A user program can read them by opening \texttt{/dev/cpu/$N$/msr}, where $N$ is the number of the CPU core, and reading it at a specific offset. Intel and AMD use different offsets for the energy-related registers and, as far as we know, the only way to determine them is to check the processor model and to read the corresponding documentation. Furthermore, the values obtained from the registers are expressed in a machine-dependent unit. Is it therefore necessary to detect the unit at runtime by decoding some other registers. Every mechanism requires a conversion step, but it is harder to implement for the MSR.

For these reasons, measuring with the MSR requires expert knowledge about the processor. We would like to insist upon the fact that implementing a distinction between AMD and Intel processors is not sufficient: while vendors often use the same offsets and units across several generations of processors, there are exceptions. For instance, Intel's manual~\cite{intelManual} gives a default value for the energy unit, but it does not apply to the \textit{dram} domain of servers equipped with Xeon E5-2600 processors.

The MSR mechanism is low-level and provides no safeguard to the developer. Reading the registers at the wrong address, or converting the bits read in the wrong way leads to a wrong measurement that can be hard to detect. Thankfully, there should be no risk of breaking the operating system nor the hardware by misusing the MSR, because we only perform reading operations, never writings.

The kernel restricts access to the MSR because it can contain sensitive information. It has been shown that the RAPL counters could be used to perform side-channel attacks, turning into a security risk~\cite{ZhangRaplSecurityLeak}. For this reason, even reading the registers requires high privileges: the binary of the measurement tool must be given the Linux capability \texttt{CAP\_SYS\_RAWIO} (or \texttt{CAP\_SYS\_ADMIN} on old kernels) or must be run as \texttt{root}. Furthermore, access control to the MSR is all-or-nothing: the \texttt{msr} module offers no way to limit a program to a restricted set of registers. This problem can be mitigated with the \texttt{msr-safe} module~\cite{msrSafe}, but the other difficulties remain.

\subsection{Power Capping Framework}
\label{subsec:powercap}

The Power Capping framework (powercap)~\cite{powercapDoc} is a software interface provided by the Linux kernel on top of the low-level RAPL interface. It allows to control RAPL from userspace through the sysfs virtual file system.
The hierarchy of the sysfs under \texttt{/sys/devices/virtual/powercap/intel-rapl} resembles the hierarchy of the RAPL domains, allowing tools to discover which domains are available on the computer.

There is one confusing difference, though: in contrast to figure~\ref{fig:raplDomains}, powercap puts the \textit{dram} domain inside of the \textit{package} domain. We have found no justification of this difference in the other works related to RAPL. We think that it can be explained by the fact that, in a multi-socket system, each CPU typically has its own directly-accessible memory. Since the registers are provided by the CPU, the \textit{dram} energy counter of each CPU is different and only reports the consumption of the memory that is attached to it. Powercap's authors would have chosen to reflect this by putting the \textit{dram} domain next to the \textit{core} domain, for each socket. Even so, our tests demonstrate that the energy consumption of the memory is not included in the energy consumption of the \textit{package}.

% An example of such a hierarchy is given by the figure~\ref{fig:powercapHierarchy}. Each sub-folder of `intel-rapl` corresponds to a domain.

% \begin{figure}
%     \centering
%     \includegraphics[width=\columnwidth]{diagrams/psys-rapl-hierarchy.png}
%     \caption{Folder hierarchy of the RAPL powercap sysfs on a recent laptop with an Intel CPU. For each subfolder, we indicate the name of the corresponding domain.}
%     \label{fig:powercapHierarchy}
% \end{figure}

% https://superuser.com/questions/916516/is-the-amount-of-numa-nodes-always-equal-to-sockets
% https://www.reddit.com/r/homelab/comments/13uv4jw/how_does_having_a_dual_socket_motherboard_affect/

Being a high-level interface, powercap is easier to use than the MSR. On one hand, the measurements are provided in text files, and the MSR unit conversion is automatically applied. This is handy for testing or developing simple scripts. On the other hand, overflows frequently occur, like with the MSR mechanism. Almost no knowledge about the processor is required, even if one must be careful with the meaning of the domains' hierarchy, as previously explained. No special case is required, because the Linux kernel takes care of them.

Powercap just needs read access on the \texttt{intel-rapl} directory. The file permissions or access control lists can be adjusted accordingly, which allows to limit the privileges of the measurement tools.

Finally, the resiliency of the mechanism is good, thanks to an automatic adaptation of the hierarchy of the sysfs to the available domains. Of course, to benefit from this advantage, the hierarchy must not be hard-coded in the measurement tool. Our reference implementation is able to adapt to the availability of the domains.

\subsection{perf-events}
\label{subsec:perfevent}

The "perf events" subsystem (hereafter "perf-events") is another Linux kernel interface. It provides event-oriented performance monitoring capabilities~\cite{perfEventPage}. There are two types of events: counting events, whose latest values must be polled, and sampling events, which are periodically added to a buffer. The RAPL energy counters have been integrated as counting events. Therefore, periodically polling the counters to detect the overflows is necessary, as with the other mechanisms. An advantage of perf-events is that overflows are already corrected by the subsystem. Of course, the value returned by perf-events has a limited size (64 bits integer), so an overflow is still possible in theory, but very unlikely in practice. Thus, the probability of reporting erroneous measurements is much lower than with MSR and powercap.

Usually, polling the events is done from userspace. First, the list of available events can be read from the sysfs, by inspecting the \texttt{/sys/devices/power/events/} directory. Each event corresponds to a RAPL domain and needs to be opened by calling \texttt{perf\_event\_open}~\cite{perfEventOpenDoc}. The obtained file descriptor can then be read periodically to access the measurements. Unlike powercap, perf-events does not organize the events in a hierarchy matching the different CPU sockets. The only difficulty that we have found is that each event must be opened exactly once for each socket. Using perf-events requires no expert knowledge and allows operating at a higher level of abstraction than the MSR mechanism.

In terms of resiliency, perf-events is less versatile than powercap, because one needs to call \texttt{perf\_event\_open}. It is therefore not available in every language, and it is much harder -- though not impossible -- to use it in Bash scripts. Yet, when used in a compatible environment, perf-events is as easy to maintain over time as powercap, and can easily be adapted to support new domains and microarchitectures.

 Fine-grained access control is possible, since it requires either the \texttt{CAP\_PERFMON} capability (\texttt{CAP\_SYS\_ADMIN} before Linux 5.8) or the kernel setting \texttt{perf\_event\_paranoid} set to $0$ or $-1$.

We have analyzed here the usual mechanism implemented on top of perf-events. Another way of handling the events is to read the file descriptors from kernel space using the recent eBPF technology, as described in the following section.
As far as we know, this technique has not been investigated yet.

\subsection{perf-event via eBPF}
\label{subsec:ebpf}

Originally, eBPF was an abbreviation for "extended Berkeley Packet Filter", but it is now a standalone name~\cite{ebpfWebsite}. It refers to a technology that allows to inject code into a kernel, in particular the Linux kernel since version 3.15~\cite{ebpfKernelVersions}. Thanks to eBPF, a user program can be attached to an existing function of the kernel, which has proven to be useful for implementing fast packet filtering~\cite{ScholzPerfEbpf} and profiling.

In this context, we want to examine whether eBPF is a good fit for measuring the energy consumption with RAPL. The idea is that, under the hood, the aforementioned mechanisms will lead to a call to the \texttt{rdmsr} x86 instruction, which needs to be done from the kernel. By reading the values with eBPF, all our measurement code would be executed by the kernel, and we would not need to switch from user mode to kernel mode. This could lower the overhead of the measurement. We test this hypothesis by designing a new measurement mechanism based on eBPF and by benchmarking its implementation (see section~\ref{sec:bench-protocol}).

\begin{figure}[h]
    \centering
    \includegraphics[width=0.90\linewidth]{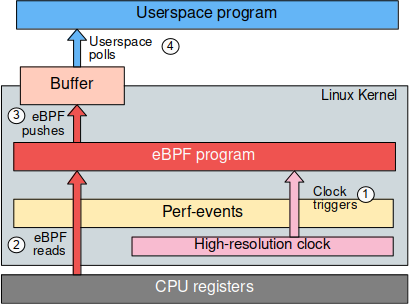}
    \caption{Measurement mechanism based on perf-events and eBPF}
    \label{fig:raplEbpf}
\end{figure}

% explain the architecture of the eBPF-based mechanism, as implemented in our open-source reference program
Figure~\ref{fig:raplEbpf} illustrates the inner workings of this new mechanism. As a preliminary step, \texttt{perf\_event\_open} is called and the returned file descriptors are passed to the eBPF program. This program is then injected into the Linux kernel and attached to a \texttt{SF\_CPU\_CLOCK} event. At a given frequency (1000 Hz in our implementation), the clock triggers the program, which reads the RAPL energy counters via perf-events and pushes the measurements into a buffer. At a regular interval managed by another timer, the userspace program polls the content of the buffer and obtains the energy consumption measurements.

Using eBPF is significantly more complicated than using the "regular" userspace variant of perf-events, that is why we evaluate its technical difficulty to "high" (see section~\ref{subsec:comparison-synthesis}).

In terms of safeguards, it is just as robust to overflows as the regular perf-events, but other types of errors can occur. For example, we hit multiple difficulties related to the transfer of values between user and kernel space. Our implementation is written in Rust, which prevents certain types of error. We believe that it would have been even more challenging to write it in C.

Using eBPF requires to have the capability \texttt{CAP\_BPF} ( \texttt{CAP\_SYS\_ADMIN} before Linux 5.8) or to be run as root.

This mechanism demands a bit more work to adapt to changes, because the sizes of the buffers must be adjusted, and the injected code must be updated in consequence. In fact, old versions of the Linux kernel are still used in production, and they do not support loops in eBPF kernel code. That is why the implementation we propose uses a match-case on the number of RAPL domains. Supporting a new domain hence requires to update the code.

\subsection{Synthesis}
\label{subsec:comparison-synthesis}

Table~\ref{table:mechanismComparison} below summarizes the results of the previous comparative analysis.

\newcommand{\cgood}[1]  {\cellcolor{green!30}{#1}}
\newcommand{\cmedium}[1]{\cellcolor{orange!30}{#1}}
\newcommand{\cbad}[1]   {\cellcolor{red!30}{#1}}

\begin{table}[h]
    \centering
    %\begin{tabular}{|c|c|c|c|c|c|}
    \begin{tabular}{|p{0.12\linewidth}|p{0.11\linewidth}|p{0.1\linewidth}|p{0.13\linewidth}|p{0.13\linewidth}|p{0.13\linewidth}|}
    \hline
        {mechanism} & technical difficulty & required knowledge & safeguards & privileges & resiliency \\
    \hline
        MSR               & \cmedium{medium} & \cbad{CPU knowledge} & \cbad{none} & \cbad{{\scriptsize SYS\_RAWIO} cap. + msr module} & \cbad{poor} \\
        \hline
        perf-events +~eBPF & \cbad{high (long, complicated code)} & \cgood{limited} & \cmedium{overflows unlikely, many other possible mistakes} & \cmedium{{\scriptsize PERFMON} and {\scriptsize BPF} capabilities} & \cmedium{manual tweaks necessary for adaptation} \\
        \hline
        perf-events & \cgood{low} & \cgood{limited} & \cgood{good, overflows unlikely} & \cmedium{{\scriptsize PERFMON} capability} & \cgood{good} \\
        \hline
        powercap          & \cgood{low} & \cgood{limited} & \cmedium{beware of overflows} & \cgood{read access to one dir} & \cgood{good, very flexible} \\
    \hline
    \end{tabular}
    \vspace{.5em}
    \caption[short]{Qualitative comparison of the measurement mechanisms}
    \label{table:mechanismComparison}
\end{table}

\section{Experimental study of the measurement overhead}
\label{sec:experiments}

% PLAN de cette section :
% voilà ce qu'on veut faire, protocole détaillé
% résultats et analyse des résultats (ne pas mettre tous les détails techniques de l'analyse statistique des résultats)
% mettre en avant les éléments inattendus s'il y en a

% NB : carbon0 = AMD !! naboo23 = Intel

\subsection{Benchmarking Protocol}
\label{sec:bench-protocol}

We describe here the protocol of the experimental study we conducted in order to assess the impact of the measurement mechanisms. This quantitative analysis completes the previous qualitative comparison. We are interested in the performance overhead on the other running applications, but also on the energy consumption of the machine when it is idle.

\subsubsection{Test Environment}

To run the benchmarks, we used two dedicated server-class machines in Atos' datacenters.
The first one was equipped with two AMD EPYC 7702 64-cores processors and RHEL (Red-Hat Enterprise Linux) 8.7.
The second was composed of two Intel Xeon E5-2680 v4 14-cores processors and ran RHEL 8.5.
Both servers were running the Linux kernel version 4.18.0.
On the AMD machine, the only domain available in every mechanism was the \textit{package} domain.
On the Intel machine, we were able to measure both \textit{package} and \textit{dram} domains.

We chose to run the well-known NAS benchmarks~\cite{nasBench95}, mainly to allow our results to be compared with other papers such as the experiments of Jay et al.~\cite{JayCcgrid}. In order get various types of parallel applications, we used three NAS benchmarks: BT (block tri-diagonal solver), CG (conjugate gradient) and EP (embarrassingly parallel). The benchmarks were compiled on each server and configured to use all the cores.

\subsubsection{Inputs and Outputs}
We recorded the running time of the NAS program, as reported by itself. Thus, this metric only includes the time it takes to solve the benchmark's problem, not the call to \texttt{posix\_spawn}, nor the final clean-up of the program's memory, etc.
We also performed some runs with a sleep of 20 minutes instead of a NAS program in order to evaluate the impact of the measurement on an idle CPU. During such a sleep, we gathered statistics about the state of the CPU with the \texttt{turbostat} command.
During the experiments, we used a script to change the following variables:
\begin{itemize}
    \item the NAS benchmark to execute (BT, CG, EP, or a 20-minutes sleep)
    \item the measurement mechanism to use (MSR, powercap, perf-events userspace, perf-events eBPF)
    \item the frequency of the measurement (0 i.e. no measurement, or 0.1Hz, 1Hz, 10Hz, 100Hz or 1000Hz)
    \item the RAPL domains to measure (\textit{package} only, or \textit{package} and \textit{dram})
\end{itemize}

That makes 96 combinations to test on the AMD server, and 192 on the Intel server. For every run of a combination, we recorded the aforementioned metrics, the variables, and the current time. At the same time, we recorded the evolution of the power consumption of each server by querying their BMC. This was done from another server, in order not to interfere with the benchmarks.

\subsubsection{Benchmark repetitions}

\begin{figure}[h]
    \centering
    \includegraphics[width=\linewidth]{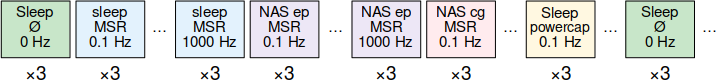}
    \caption{Benchmark repetitions spread over time}
    \label{fig:benchmarkSpread}
\end{figure}

We ran the benchmarks for approximately one month, for as long as we could reserve the servers for our experiments. Instead of running many repetitions of one combination, then many repetitions of another combination, and so on for every combination until the end, we spread the repetitions over time.
As illustrated by figure~\ref{fig:benchmarkSpread}, we ran 3 repetitions of the first combination, then 3 repetitions of the next one, etc.
Once all the combinations were done, we started all over again: 3 repetitions of the first combination (the rightmost green square on figure~\ref{fig:benchmarkSpread}), and so on.

The goal of this spreading is to avoid bias we cannot control, such as another user executing something, a technical problem in the datacenter, a change in weather that affects the temperature of the room, etc. These external factors are likely to happen at a specific point in time, and thanks to this way of benchmarking they have a lower chance to spoil the results of a specific combination.

\subsubsection{Outliers}
% publier aussi les outils de bench ? ça serait chouette -> dire qu'ils sont disponibles sur le github
% carbon0 = AMD, naboo23 = Intel !!

After the execution of the benchmarks, we analyzed them with statistical tools. The first step was to eliminate the outliers in each combination. To identify them, we used the usual interquartile range. A value was considered to be an outlier if it was outside of the range $[q_1 - 3(q_3-q_1), q_3 + 3(q_3-q_1)]$, where $q_1$ is the first quartile and $q_3$ is the third quartile.
After this step, we were left with 2186 observations for the AMD server and 2564 for the Intel one.

\subsection{Results}
\label{subsec:benchmark-results}

\subsubsection{Impact on parallel software}
\label{subsec:benchmark-results-nas}

\begin{figure}[h]
    \centering
    \includegraphics[width=\columnwidth]{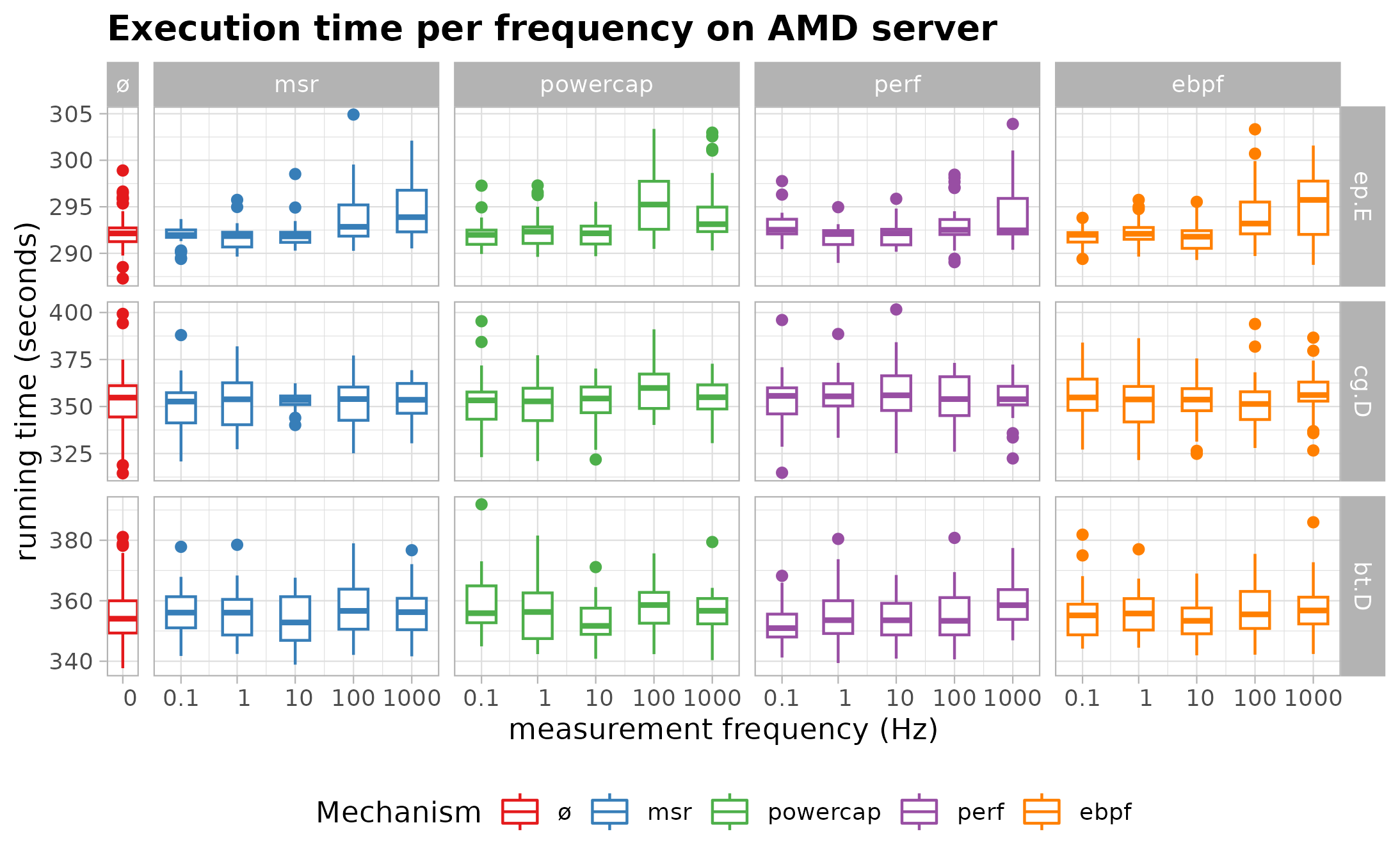}
    \caption{Performance impact of the measurement on NAS benchmarks (AMD server). The leftmost column contains the baseline.}
    \label{fig:runningTimeAMD}
\end{figure}

\begin{figure}[h]
    \centering
    \includegraphics[width=\columnwidth]{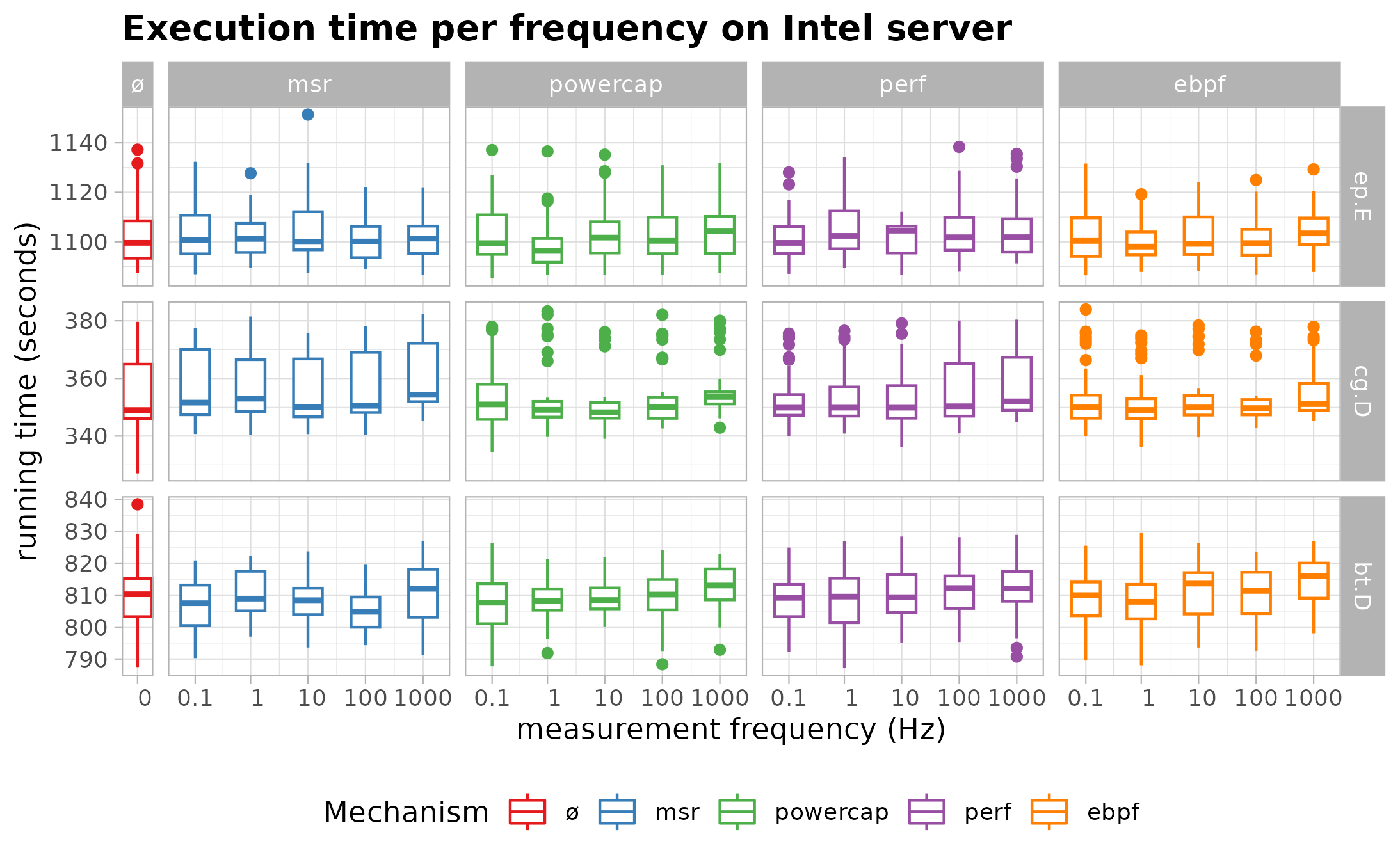}
    \caption{Performance impact of the measurement on NAS benchmarks (Intel server). The leftmost column contains the baseline.}
    \label{fig:runningTimeIntel}
\end{figure}

Figures~\ref{fig:runningTimeAMD} and~\ref{fig:runningTimeIntel} show the running time of the three NAS benchmarks on the AMD and Intel server respectively. Our goal here is to determine whether some mechanisms are more efficient than others, and whether increasing the frequency makes the potential performance overhead more visible.

\begin{figure}[h]
    \centering
    \includegraphics[width=\columnwidth]{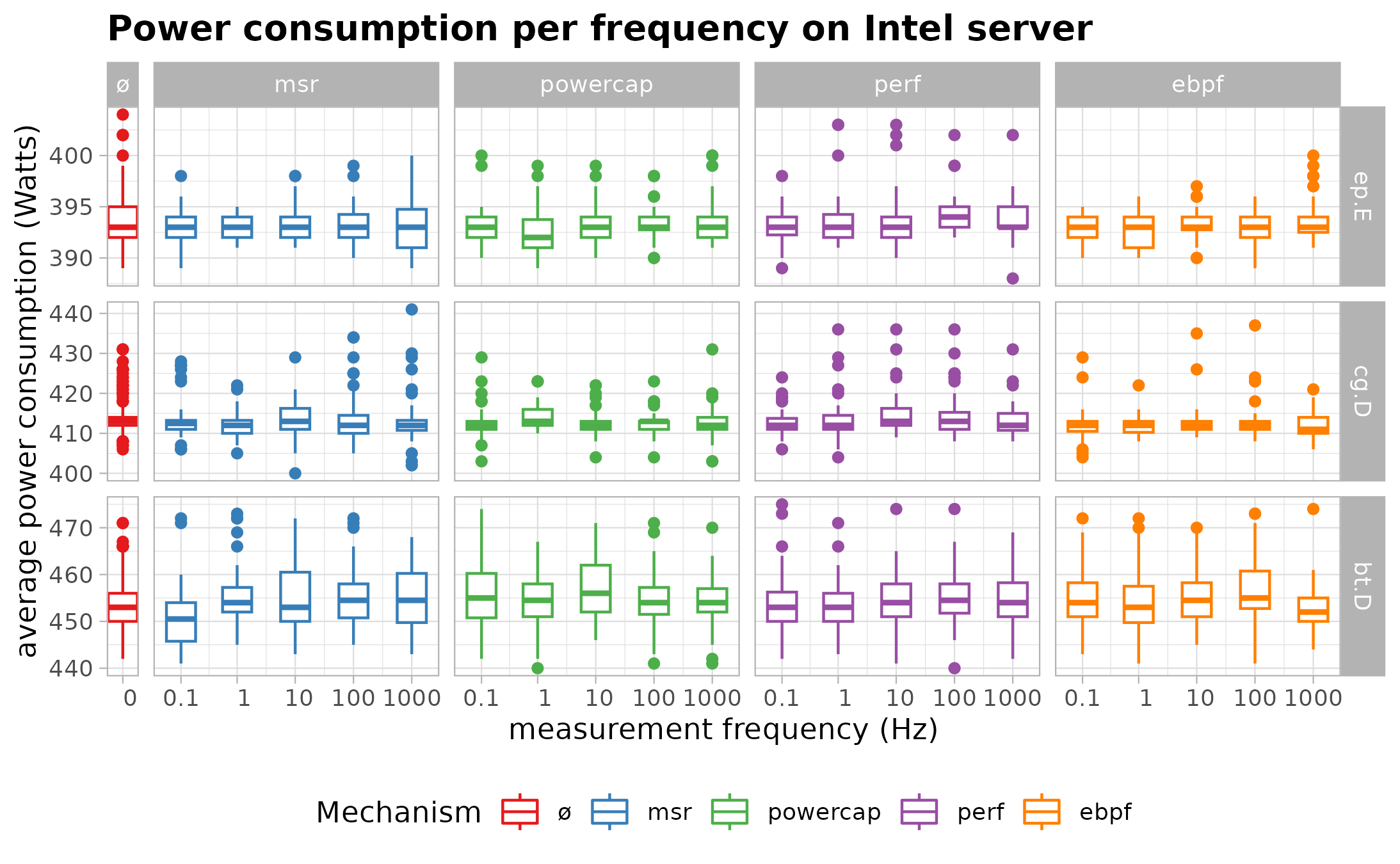}
    \caption{Impact of the measurement on power consumption during NAS benchmarks (Intel server). The leftmost column contains the baseline.}
    \label{fig:runningPowerIntel}
\end{figure}

Visually, it is difficult to see an overhead effect, even with a high measurement frequency. To test the existence of an overhead in a more rigorous way, we conducted a statistical test on each group. A group here is a subset of the data referring to the same server, NAS benchmark and RAPL mechanism. Our data did not meet the criteria for an ANOVA test (analysis of variance), therefore we applied a one-sided Wilcoxon rank sum test. We used the standard value of $5\%$ for the significance level $\alpha$ and applied the Holm-Bonferroni correction on the p-values.
For each group we compared the running time of each frequency to the running time obtained without any measurement. The latter is represented by red boxplots on the left of figures \ref{fig:runningTimeAMD}, \ref{fig:runningTimeIntel} and \ref{fig:runningPowerIntel}. Our alternative hypothesis for the statistical test is then "the running time with measurement is higher than without".

The test shows that the overhead is statistically significant on the "ep.E" benchmark on the AMD server, at a frequency of 1000 Hz for msr, powercap and eBPF, and a at frequency of 100 Hz for powercap. While significant, the overhead remains small, with a location shift estimated at 1.5 to 3.4 seconds, that is, 0.5\% to 1.2\%  (see appendix~\ref{appendix-app:A}). The location shift is an estimator of the median of the difference between a sample of the first group (with some measurement frequency) and a sample of the second group (without any measurement). Surprisingly, perf-events seems to have a smaller overhead than the other mechanisms on this benchmark. In particular, the low-level MSR interface seems slower than perf-events, which is counter-intuitive. We discuss this point in section~\ref{sec:synthesis}. Regarding the other benchmarks, there is not enough evidence to say that measuring increases the running time.

The AMD server is not impacted by the measurement on all benchmarks. Only the "ep.E" benchmark, which consist of heavy floating-point operations is slowed down. The other benchmarks, which use more memory operations, are not significantly affected. Interestingly, the Intel server does not exhibit this behavior, since the only significant differences are on "cg.D" and msr, and "bt.D" and eBPF, with a location shift of respectively 5.6 seconds (1.6\%) and 5.7 seconds (0.7\%). An acquisition frequency of 10 Hz or less causes a negligible impact on the running time in every tested configuration.

In comparison, we measured that the modified versions of Scaphandre and CodeCarbon, which have a restricted feature set as described in section~\ref{subsec:ensuringAccuracy}, have a non-negligible overhead of respectively 3-4\% (depending on the benchmarks) and 0.5-0.8\% at 10~Hz on the Intel server.

Note that figure~\ref{fig:runningTimeIntel} does not make a difference between the benchmarks ran while measuring one RAPL domain and the benchmarks ran while measuring two RAPL domains.
This is because, according to our analysis, that there is insufficient evidence to conclude in favor of a difference between these two cases. In other words, measuring one more domain does not change the overhead of the measurement on the benchmarks.

In addition to the running time, we recorded the power consumption of the entire servers, using their BMC. As can be seen on figure ~\ref{fig:runningPowerIntel}, we found that the measurement operation does not impact the average power when the machine is under heavy load. The AMD server showed results that were similar to the Intel server.

% \begin{figure}[h]
%     \centering
%     \includegraphics[width=\columnwidth]{ggplot/nas-boxplots-avg-power-carbon.png}
%     \caption{Impact of the measurement on power consumption during NAS benchmarks (AMD server). The leftmost column contains the baseline.}
%     \label{fig:runningPowerAMD}
% \end{figure}

This lack of overconsumption under heavy load can be explained by the fact that, when running the benchmarks, the OS puts the CPUs at their maximum power. Hence, it cannot increase further.

\subsubsection{Impact on idle CPU}
\label{subsec:benchmark-results-idle}

We have just showed that the impact of RAPL energy measurements on a loaded CPU was low. In this section, we show that there can be a significant impact on an idle CPU, depending on the measurement frequency. To evaluate this impact, we used the data collected during the "sleep" runs, which consist of a simple sleep of 20 minutes. In addition to the total power read with the BMC, we collected the C-States of the processor. The state "C0" indicates that the processor is active and executing instructions. Other states are idle states, during which a subset of the CPU is disabled in order to save power.

\begin{figure}[h]
    \centering
    \includegraphics[width=\columnwidth]{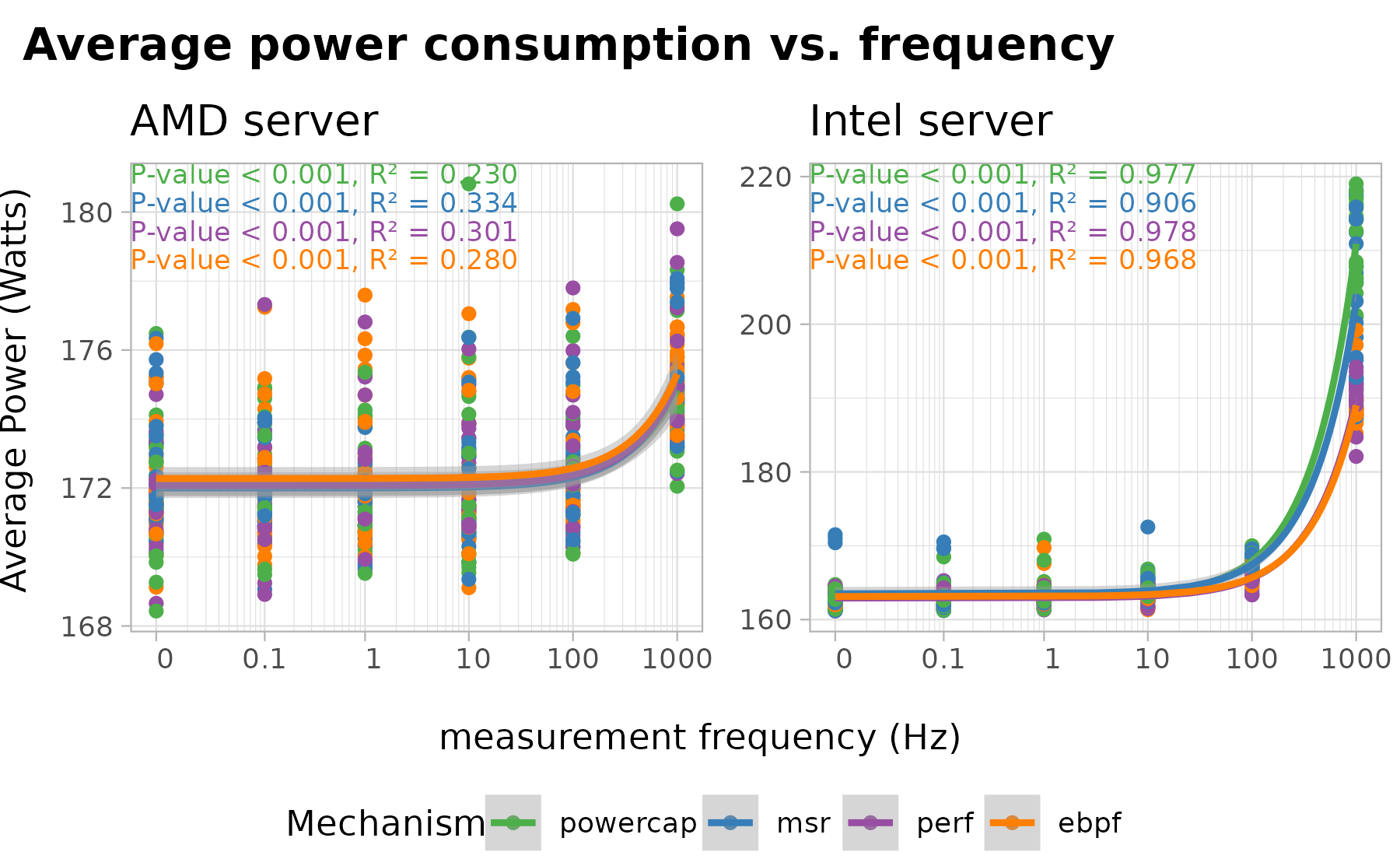}
    \caption{Relationship between acquisition frequency and idle power consumption (AMD and Intel server)}
    \label{fig:idlePower}
\end{figure}

First, let us look at the average power consumption of the server during each run of the benchmarks.
As we can see in figure~\ref{fig:idlePower}, the two servers do not react in the same way.

On the AMD server, the power increases starting at a measurement frequency of 1000 Hz, for which the machine draws around 2.7 to 3.3 Watts more than without measurement. On the Intel server, a Wilcoxon rank sum test indicates that the effect is significant starting at 10 Hz for powercap and msr, and 100 Hz for perf-events and eBPF. The details are available in appendix~\ref{appendix-app:B}. At the highest frequency, the average power increases by 48.6 Watts for powercap, 40.0 for msr, 27.2 for perf-events and 25.4 for eBPF. These last two mechanisms, both based on the perf-events interface, have a clear advantage when the Intel processor is idle.

\begin{figure}[h]
    \centering
    \includegraphics[width=\columnwidth]{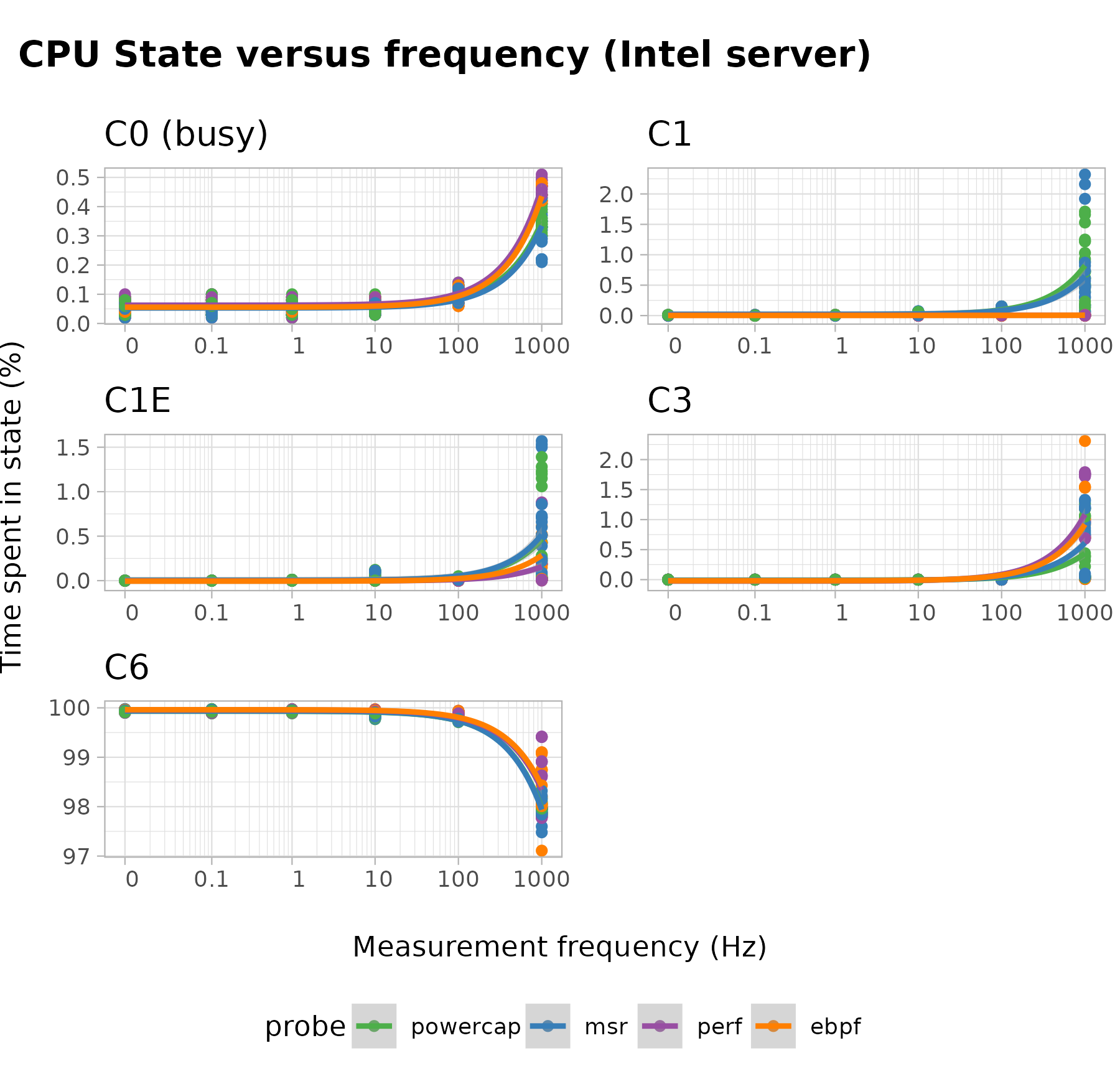}
    \caption{Relationship between acquisition frequency and idle CPU states (Intel server)}
    \label{fig:idleStateIntel}
\end{figure}

This difference between the two processors could be explained by the fact that the Intel one has more CPU states than the AMD one (note in general, Intel provides more C-States states than AMD as of 2023). As confirmed by figure~\ref{fig:idleStateIntel}, querying the RAPL counters too frequently forces the Intel CPU to spend less time in C6, which consumes the least power, and more time in the higher levels, in particular C0, which consumes the most. Increasing the measurement frequency naturally causes more instructions to be executed, hence the increase of the time spent in C0. Interestingly, a high frequency of 1000 Hz makes the CPU use the intermediary states more often (C1, C1E and C3 on the Intel CPU), to the detriment of the deepest state (C6 on the Intel CPU). This means that the system determined that it was less appropriate to use the deepest state, perhaps because of its higher exit latency (it takes more time to exit C6 than to exit C3, C1E or C1). Whether Intel's or AMD's state management is optimal is not in the scope of this work, but it is interesting to know that using high measurement frequencies while the CPU is idle forces it to reduce its use of the deepest C-States. A good strategy would therefore be to dynamically adapt the measurement frequency to the CPU load.

\begin{figure}[h]
    \centering
    \includegraphics[width=\columnwidth]{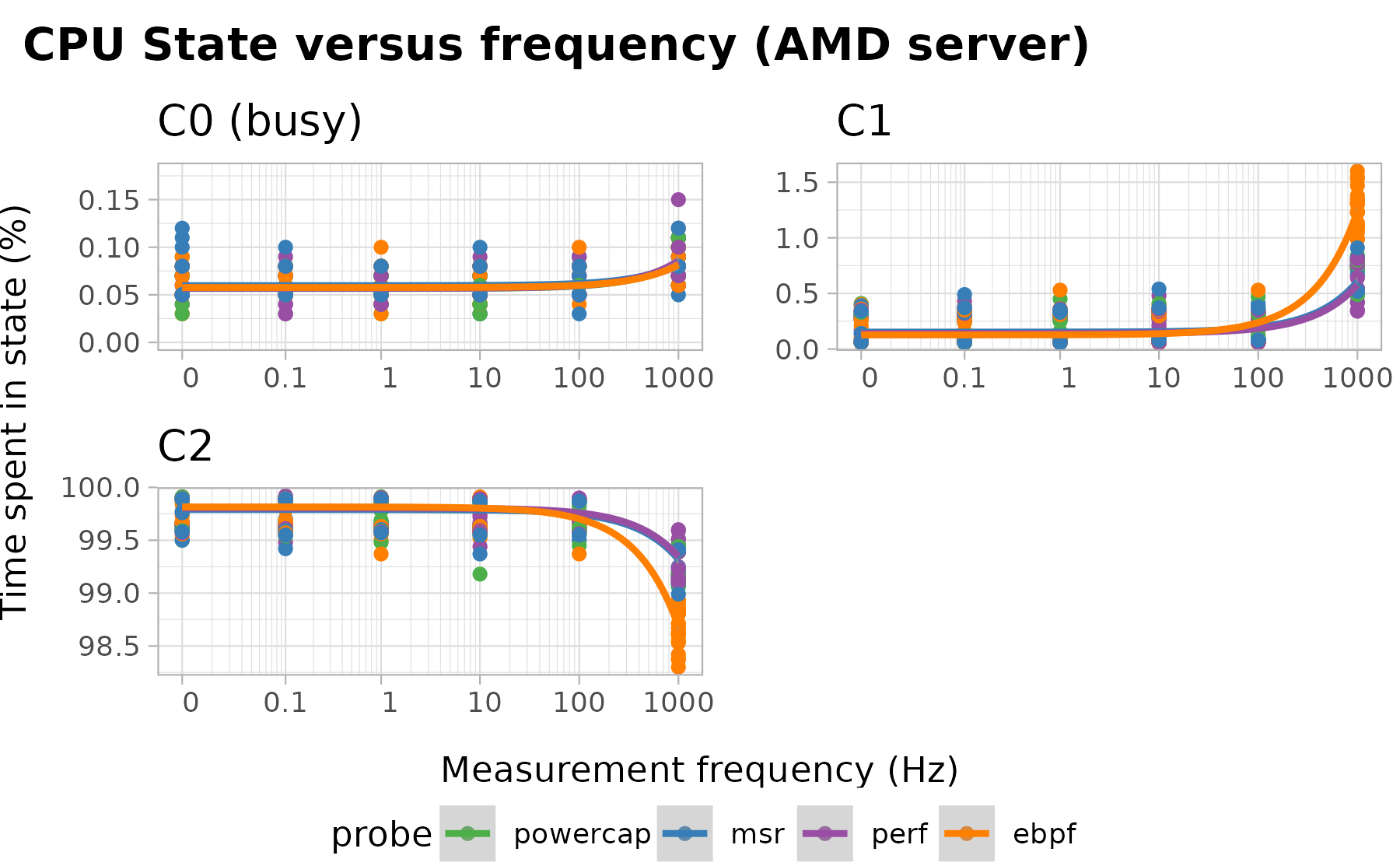}
    \caption{Relationship between acquisition frequency and idle CPU states (AMD server)}
    \label{fig:idleStateAmd}
\end{figure}

The AMD CPU also spends less time in C2 and more time in C1 and C0, as can be seen on figure~\ref{fig:idleStateAmd}. But there is obviously less differences between AMD C2 and C0 than Intel C6 and C0. The different mechanisms behave similarly with regards to the C-States.

\vspace{5pt}

\subsection{Measurement reading latency}
\label{subsec:latency}

The last experiment is a microbenchmark, that aims to assess the time it takes to read one value of a RAPL counter, using the various mechanisms. To build and run the microbenchmark, we used the Rust benchmarking tool \textit{Criterion}~\cite{criterion} on three machines: the two previously mentioned servers, and a Lenovo Thinkpad L15 Gen1 laptop running Ubuntu 22.04 LTS with Linux 6.1.

Table~\ref{table:microbench} reveals that perf-events is the fastest mechanism, followed by MSR. Not only does perf-events have the lowest latency, but it also has the lowest variance, which means that its performance is more stable than the other mechanisms. This better performance may explain the surprising result of section~\ref{subsec:benchmark-results-nas} on the AMD server. We see that powercap and MSR are faster on the laptop than on the servers. This could be due to the more recent kernel version of the laptop, i.e. these interfaces could have been optimized after Linux 4.18.

\begin{table}[h!]
    \renewcommand{\arraystretch}{1.5}
    \centering
    \begin{tabular}{|m{0.14\linewidth} | m{0.2\linewidth} | m{0.2\linewidth} | m{0.2\linewidth}|}
     \hline
     mechanism & \multicolumn{3}{m{0.6\linewidth}|}{time to read one RAPL counter \newline(95\% confidence interval, microseconds)} \\
     \hline
              & recent laptop      & AMD server     & Intel server \\
     \hline
     powercap & $[1.966, 2.106]$     & $[6.083, 6.117]$  &  $[4.230, 4.463]$ \\
     \hline
     perf-events & $[0.5371, 0.5385]$    & $[0.4990, 0.4993]$    &  $[0.7376, 0.7377]$\\
     \hline
     msr & $[0.5391, 0.5688]$            & $[4.437, 4.448]$  &  $[2.149, 2.348]$\\
     \hline
     ebpf & N/A & N/A & N/A\\
     \hline
    \end{tabular}
    \vspace{.5em}
    \caption{Table of the cost of polling one measurement from RAPL}
    \label{table:microbench}
\end{table}

As indicated by "\textit{N/A}"s in the last row, we have chosen to exclude eBPF from this microbenchmark because of its non-representative results. In the novel eBPF-based mechanism presented in section \ref{subsec:ebpf}, the reading of the counter, as most of the work, happens in the kernel, and thus cannot be measured by the Criterion benchmark tool.

\subsection{Synthesis and recommendations}
\label{sec:synthesis}

% bien structurer les différentes recommendations
Our experiments reveal that reading the RAPL energy consumption counters does not decrease the throughput of parallel software in almost all tested configurations.
%A small but significant overhead has been observed on an AMD server for the cpu-intensive NAS benchmark "embarrassingly parallel".
All four tested mechanisms have a small or negligible impact on the running time of the benchmarks, even when the acquisition frequency is high. The differences reported by the microbenchmarks are less visible in the long benchmark. Since the benchmarks used all the cores, we conclude that the cost of the context switches caused by the RAPL measurements is very small.

Compared to existing software tools, our work is significantly more lightweight. We have measured that the simplified versions of Scaphandre and CodeCarbon, described in section~\ref{subsec:ensuringAccuracy}, have a non-negligible overhead of at least 3\% and 0.5\% respectively, at only 10~Hz. According to prior work~\cite{JayCcgrid}, the full versions exhibit an even higher overhead (between 2 to 7\% at 1~Hz). Minimizing the overhead is good because the latter affects the results of the measurement.

We found that measuring at high frequencies had a significant impact on the idle consumption. In particular, it reduces the time spent by the processor in low-power states, which increases the energy consumption, especially on the tested Intel processor. We hence recommend adapting the acquisition frequency to the state of the node. Under heavy load, a high frequency can be used in order to capture more information about the running processes. On the opposite, when it is lightly loaded a lower frequency should be used. In addition, the perf-events and eBPF mechanisms seem to be the most energy-efficient when the processor is idle.

While eBPF can be used to obtain the energy consumption reported by RAPL, it has no advantage over the other mechanisms. In light of its complexity, analyzed in section~\ref{subsec:ebpf}, we recommend not to use eBPF for this task. We make the same recommendation about MSR: unless there is no choice, that is if the OS does not provide any interface on top of the registers (like Windows), the user will prefer a higher-level mechanism.

A counter-intuitive result of our experiments is that MSR is actually slower than perf-events for energy measurement. Analyzing the implementation of the MSR module may provide an explanation. As detailed in section~\ref{subsec:msr}, reading the counters is achieved by reading the file \texttt{/dev/cpu/$N$/msr}, where $N$ is the id of the CPU core. The module relies on the file's metadata to determine on which CPU core the \texttt{rdmsr} x86 instruction must be executed, and fetches some information related to the inode on every read. On the contrary, perf-events keeps track of the relevant information in a simple structure initialized on \texttt{perf\_event\_open}, not on every read.

Perf-events and powercap have similar qualities, yet important differences. On one hand, perf-events is almost immune to overflows, consumes less on idle and has a lower latency. On the other hand, powercap uses the more friendly sysfs API, which allows to get the measurements by simply reading a text file. If possible, perf-events should be preferred for its efficiency, but can be harder to use in some contexts like Bash scripts. The choice is therefore in the hands of the developer.

\section{Conclusion}
\label{conclusion}

In this work, we considered the different software mechanisms that allow to access RAPL energy counters, including a new eBPF-based mechanism that we designed. We provided a precise understanding of their operations and compared them on the basis of qualitative criteria, namely, technical difficulty, required knowledge, provided safeguards, necessary privileges and resiliency. To highlight their differences and help the developers of software measurements tools, we released a new minimal tool with a reference implementation of each studied mechanism. We explained how difficulties such as overflows, inaccurate timing and I/O jitter could be overcome. We assessed that our implementation was able to sustain high measurement frequencies with a low CPU overhead and very low standard deviation, hence improving over the two existing tools that we tested.

Using this minimal tool, we conducted an experimental study on two processor models. Our results showed that, despite their differences, the mechanisms had a similar performance and energy overhead when the machine was loaded. Under nearly all benchmark configurations, running the tool had a negligible impact on the running time and power consumption, even with a high acquisition frequency of 1000~Hz. This indicates that our implementation is more lightweight and more efficient than existing software tools. It would therefore be interesting to add more features to the minimal tool, such as the estimation of the consumption of each individual process, while keeping its overhead small. The aim would then be to extend the minimal tool in order to make it better suited for end users.

We found notable differences of overhead between the mechanisms on an idle server and in microbenchmarks. In light of the experimental results and qualitative comparison, we were able to provide recommendations on the choice of the most adequate mechanism. Our main recommendation is that, quite unexpectedly, one does not need to use the most complex mechanisms (MSR, eBPF) in order to be efficient. Prefer to use the perf-events interface whenever it is possible.

Dynamically adapting the acquisition frequency to the load is also recommended. It could be particularly useful to limit the energy consumption of edge devices while collecting precise measurements. This idea could be explored in a future work, with more diverse benchmarks.

This study did not consider GPUs nor TPUs. It looks like there are fewer choices to make since GPU vendors provide a relatively high-level software library (NVML for NVidia, ROCm for AMD) and impose an underlying mechanism. Nevertheless, a natural research extension is to conduct a similar comparative and experimental study on these specialized hardware.

\newpage

%%%%%%%%%%%%%%%%%%%%%%%%%
%\section*{Acknowledgment}

% pourquoi pas

%\section*{References}

\cleardoublepage
% \bibliography{IEEEabrv,biblio.bib}
\printbibliography

\begin{IEEEbiography}[{\includegraphics*[width=1in,clip]{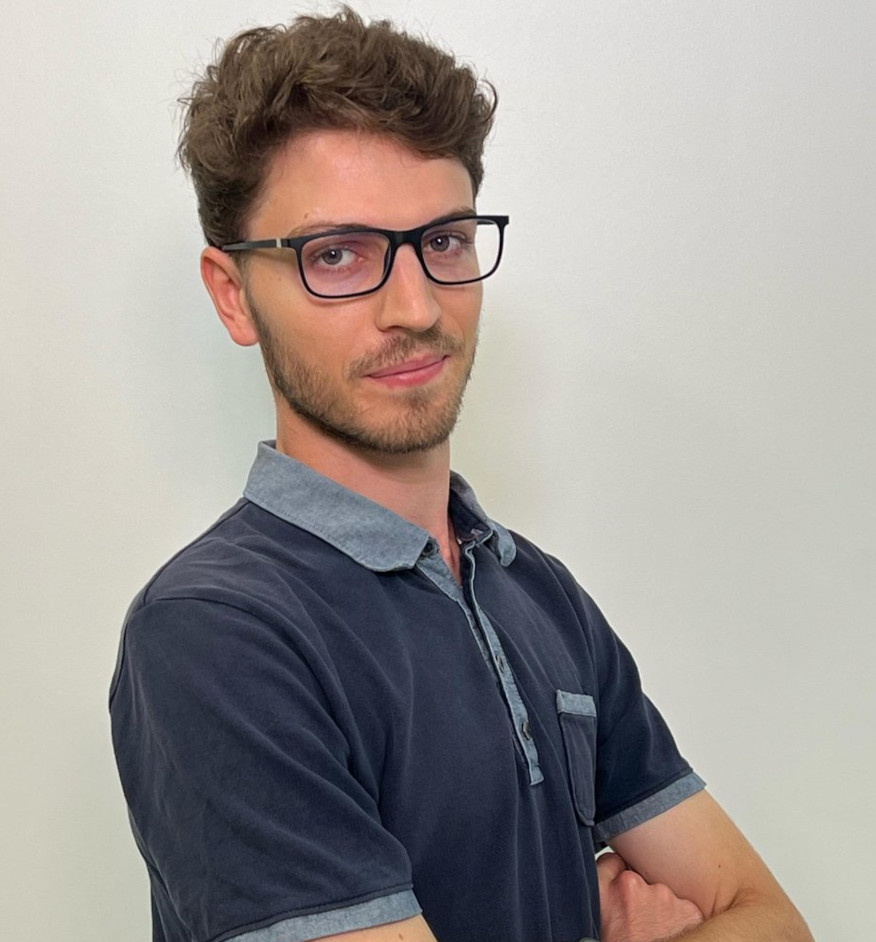}}]
    {Guillaume Raffin} is a PhD student at the institute of technology of Univ. Grenoble-Alpes and at the R\&D department of Bull SAS, which is part of Eviden (Atos group). He is working on distributed multi-site scheduling and measurement techniques, with a focus on environmental impact and industrial applications in both traditional HPC and cloud computing.
\end{IEEEbiography}

\begin{IEEEbiography}[
    {\includegraphics*[width=1in,clip]{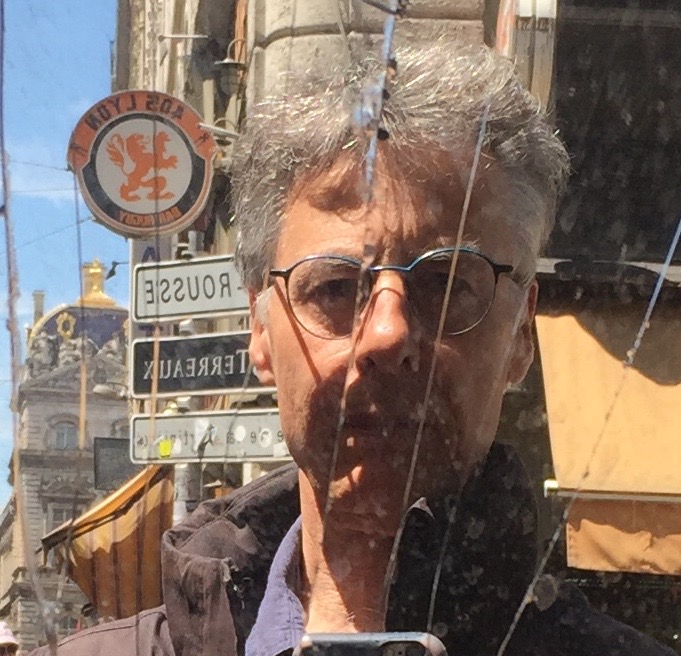}}
]{Denis Trystram} is a distinguished professor at the institute of technology of Univ. Grenoble-Alpes.
He is an honorary member of the Institut Universitaire de France.
He was working for a long time on resource management in parallel and distributed platforms (including HPC clusters, clouds, Internet of Things)
and multi-objective optimization with a special focus on minimizing the energy consumption.
Since 2019, he is leading a research program at the multidisciplinary AI institute in Grenoble on edge intelligence and frugal AI.
His academic record is composed of more than 100 publications in international peer reviewed journals and more than 150 conferences.
More details can be found at: http://datamove.imag.fr/denis.trystram/index.php
\end{IEEEbiography}

\vfill

\newpage
\appendices

\section{Results of the Wilcoxon rank-sum test on the running time of the benchmarks}
\label{appendix-app:A}

The \textit{adj.pvalue} column contains the corrected p-values. The \textit{shift} refers to the location shift estimator.

% latex table generated in R 4.3.1 by xtable 1.8-4 package
% Fri Oct  6 09:43:49 2023
\begin{table}[ht!]
    \centering
    \begin{tabular}{|llrrrr|}
        \hline
      nas\_bench & mechanism & freq & pvalue & adj.pvalue & shift \\
        \hline
      ep.E & msr & 0.10 & 0.85 & 1.00 & -0.22 \\
        ep.E & msr & 1.00 & 0.96 & 1.00 & -0.39 \\
        ep.E & msr & 10.00 & 0.88 & 1.00 & -0.29 \\
        ep.E & msr & 100.00 & $< 0.01$ & 0.22 & 1.08 \\
        ep.E & msr & 1000.00 & $< 0.01$ & $< 0.01$ & 1.94 \\
        ep.E & powercap & 0.10 & 0.80 & 1.00 & -0.20 \\
        ep.E & powercap & 1.00 & 0.25 & 1.00 & 0.18 \\
        ep.E & powercap & 10.00 & 0.53 & 1.00 & -0.02 \\
        ep.E & powercap & 100.00 & $< 0.01$ & $< 0.01$ & 2.97 \\
        ep.E & powercap & 1000.00 & $< 0.01$ & $< 0.01$ & 1.47 \\
        ep.E & perf-events & 0.10 & 0.03 & 1.00 & 0.45 \\
        ep.E & perf-events & 1.00 & 0.89 & 1.00 & -0.25 \\
        ep.E & perf-events & 10.00 & 0.54 & 1.00 & -0.02 \\
        ep.E & perf-events & 100.00 & 0.02 & 1.00 & 0.54 \\
        ep.E & perf-events & 1000.00 & 0.01 & 0.51 & 0.75 \\
        ep.E & eBPF & 0.10 & 0.91 & 1.00 & -0.26 \\
        ep.E & eBPF & 1.00 & 0.47 & 1.00 & 0.01 \\
        ep.E & eBPF & 10.00 & 0.95 & 1.00 & -0.48 \\
        ep.E & eBPF & 100.00 & $< 0.01$ & 0.10 & 1.26 \\
        ep.E & eBPF & 1000.00 & $< 0.01$ & $< 0.01$ & 3.43 \\
        cg.D & msr & 0.10 & 0.87 & 1.00 & -2.24 \\
        cg.D & msr & 1.00 & 0.44 & 1.00 & 0.50 \\
        cg.D & msr & 10.00 & 0.76 & 1.00 & -1.18 \\
        cg.D & msr & 100.00 & 0.65 & 1.00 & -0.77 \\
        cg.D & msr & 1000.00 & 0.63 & 1.00 & -0.78 \\
        cg.D & powercap & 0.10 & 0.79 & 1.00 & -1.65 \\
        cg.D & powercap & 1.00 & 0.77 & 1.00 & -2.01 \\
        cg.D & powercap & 10.00 & 0.47 & 1.00 & 0.16 \\
        cg.D & powercap & 100.00 & 0.02 & 0.97 & 5.69 \\
        cg.D & powercap & 1000.00 & 0.39 & 1.00 & 0.45 \\
        cg.D & perf-events & 0.10 & 0.47 & 1.00 & 0.21 \\
        cg.D & perf-events & 1.00 & 0.28 & 1.00 & 1.22 \\
        cg.D & perf-events & 10.00 & 0.21 & 1.00 & 2.43 \\
        cg.D & perf-events & 100.00 & 0.45 & 1.00 & 0.55 \\
        cg.D & perf-events & 1000.00 & 0.39 & 1.00 & 0.72 \\
        cg.D & eBPF & 0.10 & 0.24 & 1.00 & 1.74 \\
        cg.D & eBPF & 1.00 & 0.62 & 1.00 & -0.73 \\
        cg.D & eBPF & 10.00 & 0.66 & 1.00 & -0.88 \\
        cg.D & eBPF & 100.00 & 0.90 & 1.00 & -3.06 \\
        cg.D & eBPF & 1000.00 & 0.12 & 1.00 & 2.61 \\
        bt.D & msr & 0.10 & 0.22 & 1.00 & 1.37 \\
        bt.D & msr & 1.00 & 0.33 & 1.00 & 0.71 \\
        bt.D & msr & 10.00 & 0.77 & 1.00 & -1.45 \\
        bt.D & msr & 100.00 & 0.07 & 1.00 & 2.75 \\
        bt.D & msr & 1000.00 & 0.10 & 1.00 & 2.21 \\
        bt.D & powercap & 0.10 & 0.04 & 1.00 & 3.02 \\
        bt.D & powercap & 1.00 & 0.36 & 1.00 & 0.83 \\
        bt.D & powercap & 10.00 & 0.86 & 1.00 & -1.52 \\
        bt.D & powercap & 100.00 & 0.04 & 1.00 & 3.04 \\
        bt.D & powercap & 1000.00 & 0.17 & 1.00 & 1.65 \\
        bt.D & perf-events & 0.10 & 0.96 & 1.00 & -2.58 \\
        bt.D & perf-events & 1.00 & 0.52 & 1.00 & -0.09 \\
        bt.D & perf-events & 10.00 & 0.67 & 1.00 & -0.73 \\
        bt.D & perf-events & 100.00 & 0.54 & 1.00 & -0.17 \\
        bt.D & perf-events & 1000.00 & 0.01 & 0.30 & 4.05 \\
        bt.D & eBPF & 0.10 & 0.41 & 1.00 & 0.32 \\
        bt.D & eBPF & 1.00 & 0.24 & 1.00 & 1.10 \\
        bt.D & eBPF & 10.00 & 0.69 & 1.00 & -0.76 \\
        bt.D & eBPF & 100.00 & 0.14 & 1.00 & 1.88 \\
        bt.D & eBPF & 1000.00 & 0.06 & 1.00 & 2.61 \\
         \hline
      \end{tabular}
    \caption{Wilcoxon rank-sum tests comparing the running time with a measurement at a given frequency to the running time of the benchmarks without any measurement (AMD server).}
\end{table}

% latex table generated in R 4.3.1 by xtable 1.8-4 package
% Fri Oct  6 09:43:49 2023
\begin{table}[h!]
    \centering
    \begin{tabular}{|llrrrr|}
        \hline
      nas\_bench & mechanism & freq & pvalue & adj.pvalue & shift \\
        \hline
      ep.E & msr & 0.10 & 0.19 & 1.00 & 1.59 \\
        ep.E & msr & 1.00 & 0.17 & 1.00 & 1.67 \\
        ep.E & msr & 10.00 & 0.12 & 1.00 & 2.14 \\
        ep.E & msr & 100.00 & 0.51 & 1.00 & -0.03 \\
        ep.E & msr & 1000.00 & 0.26 & 1.00 & 1.13 \\
        ep.E & powercap & 0.10 & 0.30 & 1.00 & 0.87 \\
        ep.E & powercap & 1.00 & 0.94 & 1.00 & -2.56 \\
        ep.E & powercap & 10.00 & 0.13 & 1.00 & 2.09 \\
        ep.E & powercap & 100.00 & 0.17 & 1.00 & 1.67 \\
        ep.E & powercap & 1000.00 & 0.04 & 1.00 & 3.24 \\
        ep.E & perf-events & 0.10 & 0.32 & 1.00 & 0.77 \\
        ep.E & perf-events & 1.00 & 0.02 & 1.00 & 3.75 \\
        ep.E & perf-events & 10.00 & 0.28 & 1.00 & 0.96 \\
        ep.E & perf-events & 100.00 & 0.12 & 1.00 & 2.18 \\
        ep.E & perf-events & 1000.00 & 0.05 & 1.00 & 3.00 \\
        ep.E & eBPF & 0.10 & 0.15 & 1.00 & 1.98 \\
        ep.E & eBPF & 1.00 & 0.68 & 1.00 & -0.68 \\
        ep.E & eBPF & 10.00 & 0.31 & 1.00 & 0.79 \\
        ep.E & eBPF & 100.00 & 0.64 & 1.00 & -0.56 \\
        ep.E & eBPF & 1000.00 & 0.02 & 0.94 & 3.94 \\
        cg.D & msr & 0.10 & 0.07 & 1.00 & 1.99 \\
        cg.D & msr & 1.00 & 0.04 & 1.00 & 2.66 \\
        cg.D & msr & 10.00 & 0.31 & 1.00 & 0.73 \\
        cg.D & msr & 100.00 & 0.07 & 1.00 & 1.60 \\
        cg.D & msr & 1000.00 & $< 0.01$ & $< 0.01$ & 5.58 \\
        cg.D & powercap & 0.10 & 0.35 & 1.00 & 0.56 \\
        cg.D & powercap & 1.00 & 0.48 & 1.00 & 0.07 \\
        cg.D & powercap & 10.00 & 0.80 & 1.00 & -0.97 \\
        cg.D & powercap & 100.00 & 0.35 & 1.00 & 0.52 \\
        cg.D & powercap & 1000.00 & $< 0.01$ & 0.09 & 4.17 \\
        cg.D & perf-events & 0.10 & 0.38 & 1.00 & 0.41 \\
        cg.D & perf-events & 1.00 & 0.39 & 1.00 & 0.34 \\
        cg.D & perf-events & 10.00 & 0.53 & 1.00 & -0.16 \\
        cg.D & perf-events & 100.00 & 0.19 & 1.00 & 1.09 \\
        cg.D & perf-events & 1000.00 & 0.01 & 0.42 & 3.07 \\
        cg.D & eBPF & 0.10 & 0.40 & 1.00 & 0.39 \\
        cg.D & eBPF & 1.00 & 0.69 & 1.00 & -0.60 \\
        cg.D & eBPF & 10.00 & 0.36 & 1.00 & 0.44 \\
        cg.D & eBPF & 100.00 & 0.31 & 1.00 & 0.49 \\
        cg.D & eBPF & 1000.00 & 0.03 & 1.00 & 1.97 \\
        bt.D & msr & 0.10 & 0.93 & 1.00 & -2.48 \\
        bt.D & msr & 1.00 & 0.28 & 1.00 & 0.90 \\
        bt.D & msr & 10.00 & 0.83 & 1.00 & -1.38 \\
        bt.D & msr & 100.00 & 1.00 & 1.00 & -4.38 \\
        bt.D & msr & 1000.00 & 0.16 & 1.00 & 1.94 \\
        bt.D & powercap & 0.10 & 0.82 & 1.00 & -1.64 \\
        bt.D & powercap & 1.00 & 0.81 & 1.00 & -1.56 \\
        bt.D & powercap & 10.00 & 0.58 & 1.00 & -0.27 \\
        bt.D & powercap & 100.00 & 0.41 & 1.00 & 0.35 \\
        bt.D & powercap & 1000.00 & 0.01 & 0.82 & 3.21 \\
        bt.D & perf-events & 0.10 & 0.69 & 1.00 & -0.76 \\
        bt.D & perf-events & 1.00 & 0.67 & 1.00 & -0.87 \\
        bt.D & perf-events & 10.00 & 0.28 & 1.00 & 0.87 \\
        bt.D & perf-events & 100.00 & 0.12 & 1.00 & 1.89 \\
        bt.D & perf-events & 1000.00 & 0.03 & 1.00 & 3.11 \\
        bt.D & eBPF & 0.10 & 0.53 & 1.00 & -0.12 \\
        bt.D & eBPF & 1.00 & 0.84 & 1.00 & -1.58 \\
        bt.D & eBPF & 10.00 & 0.08 & 1.00 & 2.10 \\
        bt.D & eBPF & 100.00 & 0.17 & 1.00 & 1.65 \\
        bt.D & eBPF & 1000.00 & $< 0.01$ & 0.02 & 5.71 \\
         \hline
      \end{tabular}
      \vspace{0.1cm}
      \caption{Wilcoxon rank-sum tests comparing the running time with a measurement at a given frequency to the running time of the benchmarks without any measurement (Intel server).}
\end{table}

\newpage

\section{Results of the Wilcoxon rank-sum test on the idle power consumption}
\label{appendix-app:B}

% latex table generated in R 4.3.1 by xtable 1.8-4 package
% Fri Oct  6 11:53:21 2023
\begin{table}[ht]
    \centering
    \begin{tabular}{|rlrrrr|}
      \hline
     & mechanism & freq & pvalue & adj.pvalue & shift \\
      \hline
    1 & powercap & 0.10 & 0.45 & 1.00 & 0.04 \\
      2 & powercap & 1.00 & 0.64 & 1.00 & -0.13 \\
      3 & powercap & 10.00 & 0.75 & 1.00 & -0.24 \\
      4 & powercap & 100.00 & 0.26 & 1.00 & 0.24 \\
      5 & powercap & 1000.00 & $< 0.01$ & $< 0.01$ & 2.69 \\
      6 & msr & 0.10 & 0.51 & 1.00 & -0.01 \\
      7 & msr & 1.00 & 0.91 & 1.00 & -0.44 \\
      8 & msr & 10.00 & 0.73 & 1.00 & -0.23 \\
      9 & msr & 100.00 & 0.17 & 1.00 & 0.30 \\
      10 & msr & 1000.00 & $< 0.01$ & $< 0.01$ & 2.88 \\
      11 & perf-events & 0.10 & 0.91 & 1.00 & -0.48 \\
      12 & perf-events & 1.00 & 0.61 & 1.00 & -0.11 \\
      13 & perf-events & 10.00 & 0.35 & 1.00 & 0.13 \\
      14 & perf-events & 100.00 & 0.10 & 1.00 & 0.46 \\
      15 & perf-events & 1000.00 & $< 0.01$ & $< 0.01$ & 2.77 \\
      16 & eBPF & 0.10 & 0.51 & 1.00 & -0.01 \\
      17 & eBPF & 1.00 & 0.17 & 1.00 & 0.35 \\
      18 & eBPF & 10.00 & 0.51 & 1.00 & -0.01 \\
      19 & eBPF & 100.00 & 0.61 & 1.00 & -0.08 \\
      20 & eBPF & 1000.00 & $< 0.01$ & $< 0.01$ & 3.31 \\
       \hline
    \end{tabular}
    \vspace{0.1em}
    \caption{Detailed results of the Wilcoxon rank-sum tests that compares the power of the machine with a measurement at a given frequency to the power without any measurement (AMD server).}
\end{table}

    % latex table generated in R 4.3.1 by xtable 1.8-4 package
% Fri Oct  6 11:53:21 2023
\begin{table}[ht]
    \centering
    \begin{tabular}{|rlrrrr|}
      \hline
     & mechanism & freq & pvalue & adj.pvalue & shift \\
      \hline
    1 & powercap & 0.10 & 0.63 & 1.00 & -0.06 \\
      2 & powercap & 1.00 & 0.14 & 0.95 & 0.27 \\
      3 & powercap & 10.00 & $< 0.01$ & $< 0.01$ & 1.44 \\
      4 & powercap & 100.00 & $< 0.01$ & $< 0.01$ & 4.63 \\
      5 & powercap & 1000.00 & $< 0.01$ & $< 0.01$ & 48.61 \\
      6 & msr & 0.10 & 0.67 & 1.00 & -0.15 \\
      7 & msr & 1.00 & 0.73 & 1.00 & -0.13 \\
      8 & msr & 10.00 & $< 0.01$ & $< 0.01$ & 1.64 \\
      9 & msr & 100.00 & $< 0.01$ & $< 0.01$ & 4.24 \\
      10 & msr & 1000.00 & $< 0.01$ & $< 0.01$ & 39.96 \\
      11 & perf-events & 0.10 & 0.14 & 0.95 & 0.27 \\
      12 & perf-events & 1.00 & 0.45 & 1.00 & 0.03 \\
      13 & perf-events & 10.00 & 0.06 & 0.56 & 0.35 \\
      14 & perf-events & 100.00 & $< 0.01$ & $< 0.01$ & 2.54 \\
      15 & perf-events & 1000.00 & $< 0.01$ & $< 0.01$ & 27.23 \\
      16 & eBPF & 0.10 & 0.33 & 1.00 & 0.09 \\
      17 & eBPF & 1.00 & 0.11 & 0.92 & 0.31 \\
      18 & eBPF & 10.00 & 0.05 & 0.47 & 0.44 \\
      19 & eBPF & 100.00 & $< 0.01$ & $< 0.01$ & 3.10 \\
      20 & eBPF & 1000.00 & $< 0.01$ & $< 0.01$ & 25.42 \\
       \hline
    \end{tabular}
    \vspace{0.1em}
    \caption{Detailed results of the Wilcoxon rank-sum tests that compares the power of the machine with a measurement at a given frequency to the power without any measurement (Intel server).}
\end{table}

\end{document}